\documentstyle[12pt]{article}
\input psfig
\newcommand{\la}[1]{\label{#1}}

\newlength{\numlen}

\newlength{\indexlength}

\newcommand{\be}{\begin{equation}}
\newcommand{\ee}{\end{equation}}
\newcommand{\ba}{\begin{eqnarray}}
\newcommand{\ea}{\end{eqnarray}}

\newcommand{\rmi}[1]{{\mbox{\scriptsize #1}}}

\newcommand{\tr}{\mbox{Tr\,}}

\newcommand{\bfk}{\mbox{\bf k}}
\newcommand{\bfp}{\mbox{\bf p}}
\newcommand{\bfq}{\mbox{\bf q}}
\newcommand{\bfx}{\mbox{\bf x}}

\newcommand{\fr}[2]{{\frac{#1}{#2}}}

\begin{document}

\begin{titlepage}
\begin{flushright}
CERN-TH.6973/94\\
IUHET-273\\
\end{flushright}
\begin{centering}
\vfill

{\bf 3D PHYSICS AND THE ELECTROWEAK PHASE TRANSITION: PERTURBATION
THEORY}
\vspace{1cm}

K. Farakos$^a$\footnote{Partially supported by a CEC Science program
(SCI-CT91-0729).}, K. Kajantie$^{b}$, K. Rummukainen$^c$ and M.
Shaposhnikov$^d$\footnote{On leave of absence from Institute for
Nuclear Research of Russian Academy of Sciences, Moscow 117312,
Russia.} \\

\vspace{1cm}
{\em $^a$National Technical University of Athens, Physics
Department,\\ Zografou Campus, GR 157 80, Athens, Greece\\}
\vspace{0.3cm}
{\em $^b$Department of Theoretical Physics,
P.O.Box 9, 00014 University of Helsinki, Finland\\}
\vspace{0.3cm}
{\em $^c$Indiana University, Department of Physics,
Swain Hall West 117, Bloomington IN 47405 USA\\}
\vspace{0.3cm}
{\em $^d$Theory Division, CERN,\\ CH-1211 Geneva 23, Switzerland}
\vspace{2cm}

{\bf Abstract}

\end{centering}

\vspace{0.3cm}\noindent
We develop a  method for the construction of the effective potential
at high temperatures based on the effective field theory approach and
renormalization group. It allows one to sum up the leading logarithms
in all orders of perturbation theory. The method  reproduces the
known one-loop and two-loop results in a very simple and economic way
and clarifies the issue of the convergence of the perturbation
theory. We also discuss the assumptions being made for the
determination of the critical temperature of the electroweak phase
transition, and analyse different perturbative uncertainties in its
determination. These results are then used for the non-perturbative
lattice Monte Carlo simulations of the EW phase transition in
forthcoming
paper.

\vskip 0.5in
\noindent CERN-TH.6973/94\\
\noindent IUHET-273\\
\noindent March 1994
\addtocounter{page}{1}
\thispagestyle{empty}
\pagebreak

\end{titlepage}
\section{Introduction}

The order of high temperature phase transitions in gauge theories
\cite{kir} plays an important role in cosmology. For example, first
order phase transitions in grand unified theories can drive an
inflation \cite{inf}; a first order phase transition in the standard
electroweak theory may be responsible for baryogenesis
\cite{krs,s:higgs}.

As is well known, perturbation theory in gauge theories at high
temperatures is intrinsically unreliable due to power-like infrared
divergencies \cite{linde,gpy}. In particular, the $g^6 T^4$
correction to the free energy, originating first on the 4-loop level,
is in principle not computable by perturbative methods. These
divergencies make the perturbative effective potential divergent at
small $\phi$. A logarithmic divergence appears at the 4-loop level
and at higher loops one gets power singularities\footnote{The very
high order of the ordinary perturbation theory where this happens
often leads to conclusion that these corrections are not important.
However, partial resummation of corrections with the help of exact
evolution equations \cite{wetterich} reveals that the effective gauge
coupling becomes large at $Q^2 \sim (g^2 T/\pi)^2$ making the
naive perturbative analysis at these momentum transfers
misleading.}. At non-zero $\phi$ the infrared problem is
absent, since integrals at small momenta are cut by non-zero gauge
boson masses.

In view of the above one expects that at sufficiently large $\phi$
perturbation theory is applicable. Therefore, perturbation theory
seems to be quite suitable to answer the following question: "what is
the derivative of the effective potential at {\em given} temperature
and {\em sufficiently large} $\phi$?" Note that the absolute value of
the potential does not have any physical meaning due to arbitrariness
in the overall normalization (in the absence of gravitation). In
particular, the expectation value of $\phi$ can be determined with
the help of a perturbative effective potential provided that $\phi$
is large enough. However, the determination of the {\em critical
temperature} of the phase transition, as well as determination of the
{\em bubble nucleation temperature} in cosmology is, strictly
speaking, beyond the scope of perturbation theory. Indeed, to find
the critical temperature one must know not only the value of the
potential in the broken minimum, but the value of the potential in
the phase with $\phi = 0$. While the first quantity {\em can} be
computed in perturbation theory, the second {\em cannot}. The same
remark applies also to the bubble nucleation temperature, since the
computation of the rate of the bubble nucleation requires the
knowledge of the effective action for small enough fields $\phi$. The
determination of these temperatures is impossible in perturbation
theory without {\em additional assumptions} on the behaviour of the
potential near the origin. These assumptions are to be checked by
non-perturbative methods, say, by lattice Monte Carlo simulations.

Quite an extensive literature exists on the study of the electroweak
phase transition (see, e.g. \cite{linde1}-\cite{ms}). A number of
papers were devoted to the study of daisy and superdaisy wheel
approximations for the effective potential at high temperature, where
an attempt to sum up the leading infrared divergencies was made
\cite{bbh}-\cite{amelino}. The validity of this approximation has
been discussed in \cite{ae,amelino}. A number of important
2-loop contributions to the effective potential have been
found in \cite{bd,ae}. The large magnitude of these corrections has
lead the authors to conclude that the loop expansion for the
effective potential is unreliable even for small enough Higgs masses
(say, $35$ GeV) \cite{bd,ay}. An attempt was made in \cite{ay} to
overcome this difficulty with the use of the $\epsilon$-expansion.

This paper is the first of two papers devoted to the study of the
electroweak phase transition by perturbative and non-perturbative
methods. In the present work we apply the  method based on the
effective
field theory approach \cite{weinberg} to the construction of
the effective potential at high temperatures. This method reproduces
the
known 1-loop resummed \cite{carrington,dlhll} and 2-loop resummed
results
\cite{ae}-\cite{hebecker} in a very simple and economic way. We show
that the conclusion made in refs.\cite{bd,ae,ay} on the bad behaviour
of the loop expansion due to the presence of large logarithmic
corrections is connected with the straightforward use of the minimal
subtraction scheme in 4d theory at high temperatures. We point out
how
to use the renormalization group in 3d to sum up leading logarithms
of scalar fields in all orders of perturbation theory and clarify the
issue of convergence of perturbation theory. We also discuss
the assumptions one can make for the determination of the critical
temperature of the EW phase transition, and analyse different
perturbative uncertainties in the determination of the critical
temperature.

The second paper is devoted to the non-perturbative study of the
electroweak phase transition by lattice Monte Carlo methods. We
improve there the analysis of \cite{kari} in two ways: by going from
the 1-loop to the 2-loop level in the discussion of the 3d effective
theory and its effective potential and by extending considerably the
numerical calculations of \cite{kari}. We find at small $\phi$
considerable deviations from perturbative computations of the
effective potential, indicating that the usual assumptions made for
the behaviour of the potential near the origin are wrong. We analyse
the lattice data in detail and develop further the picture of the
electroweak phase transition suggested in \cite{ms}. Given the
information provided by lattice simulations we re-analyze the
question of the Higgs mass necessary for the electroweak baryogenesis
\cite{s:higgs,ms}.

The paper is organized as follows. In section 1 we discuss the
advantage of consideration of 3d effective theories for the study of
the electroweak phase transition. Section 2 is devoted to the study
of renormalization in 3d gauge theories. We present our computation
of the 2-loop effective potential in Section 3. In Section 4 we
formulate the renormalization group approach to the 3d effective
potential. In Section 5 we consider further reduction of 3d gauge
theory and relate the theory with the $A_0$ field to the theory
without it. In Section 6 we present 2-loop formulas relating a 4d
high temperature theory to a 3d one. Section 7 contains estimates of
the parameters of the electroweak phase transition from
perturbation theory and discussion of uncertainties. Section 8 is
conclusion. Many technical details are discussed in Appendices.

\section{Dimensional reduction, dimensional regularization and
effective field theories}

Electroweak theory at high temperatures contains a number of
different mass scales. The largest one is the temperature $T$ itself,
then come the Debye screening mass, $m_D \approx g T$ and the
$\phi$-dependent W  mass, $m_W(\phi) = g \phi/2$. For the range
of temperatures relevant for cosmology the following hierarchy of
scales holds for small $g$:
\be
T \gg m_D,~m_W(\phi).
\ee

Suppose now that we compute the effective potential for the scalar
field in the minimal subtraction scheme\footnote{Everywhere in what
follows we use the modified minimal subtraction scheme
$\overline{MS}$. The scale parameter $\bar{\mu}$ in this scheme is
related to $\mu$ in dimensional regularization through $\bar{\mu}^2 =
4 \pi e^{-\gamma}\mu^2$. In order to simplify the notation we use
everywhere $\mu$ without a bar but meaning the $\overline{MS}$
scheme.}. The well known properties of this scheme are that all
counterterms in it are polynomials in mass parameters, and that
renormalization group equations are mass independent. These
properties of the MS scheme make {\em direct} computations with it in
theories with many different mass scales practically useless: one has
the appearance of logarithms of different scales, say $\log
(\mu/m_\rmi{small})$ and $\log(\mu/m_\rmi{large})$, where $\mu$ is
the unique dimensional parameter in the MS scheme. The
renormalization group prescription to take $\mu$ to be the typical
scale in the problem does not work, since it is impossible to kill
simultaneously all potentially large logarithms. This problem with
the MS scheme is well known in the discussion of grand unified
theories, heavy quark thresholds, etc. For example, the direct
computation of the cross section of the process $e^+ e^- \rightarrow$
hadrons in the MS scheme in SU(5) theory at small momenta will
immediately lead to the conclusion that the loop expansion does not
work due to the presence of large logarithms.

The way out of this problem is well known. One has to use another
subtraction scheme, in which heavy particles are decoupled (like the
MOM scheme) or use the effective field theory approach in combination
with the MS scheme \cite{weinberg}.

Given the computational advantages of dimensional
regularization, we choose the second way. The main idea of the
effective field theory approach is the construction of the Lagrangian
for light fields which is valid at small momentum transfer by
integrating out heavy particle modes \cite{weinberg}. The effective
field theory is not renormalizable -- though it is finite
if one keeps all counterterms of the original theory -- but all
non-renormalizable terms in the effective lagrangian are suppressed
by powers of the large scale. If one is not interested in power
corrections of the type $(m_\rmi{small}/m_\rmi{large})^2$, the
non-renormalizable terms can be omitted, and the effective lagrangian
describes a renormalizable theory. The coupling constants and other
parameters like particle masses are to be fixed at the large scale by
direct computation in the underlying theory. The effective theory
does not contain the heavy scale at all and, therefore, large
logarithms of the heavy scale do not appear.

Field theory at finite temperatures provides an example of the theory
where integration out of the heavy modes can be very helpful. The
euclidean time interval is finite, $0 < \tau < \beta =1/T$. This
allows one to consider 4d field theory at finite temperature
as a 3d field theory at zero temperature with an infinite number of
excitations with 3d masses $(2\pi n T)^2 + m^2$ for bosons and
$((2n+1)\pi T)^2 + m^2$ for fermions. Then, one can integrate out all
fermionic degrees of freedom (they are never massless) and all
bosonic modes with $n \neq 0$. This is the dimensional reduction of
hot 4d theories \cite{appelqpisarski}-\cite{reisz2}.

In order to escape unnecessary complications of the equations we
choose to work not in the standard model, but in SU(2) gauge theory
with a doublet of scalar fields. This theory has the same problems as
the electroweak theory in the infrared region. At the same time,
incorporation of fermions does not cause any problems, while the
inclusion of the U(1) factor is straightforward but leads to lengthy
formulas (see, e.g. \cite{carrington,ae}).

Just to fix the notations, the 4d lagrangian of the theory
under consideration is
\be
L={1\over4} F_{\mu\nu}^aF_{\mu\nu}^a +
(D_{\mu}\phi)^\dagger(D_{\mu}\phi) -\fr12 m^2\phi^\dagger\phi
+\lambda(\phi^\dagger\phi)^2.
\label{4dlagr}
\ee
We were not able to find in the literature a complete 1-loop analysis
of dimensional reduction in this theory. Partial results are
contained in the paper by Landsman \cite{land} who has made a
computation in the pure SU(N) gauge theory\footnote{Some of the
equations in the appendix of \cite{land} must be taken with care. For
example, eq.(A.7), as it appears in that paper is not correct. We
suspect
that this is due to a misprint, since we checked a number of terms in
the final equation (6.3) with the use of correct version of (A.7),
and found a result coinciding with (6.3) of \cite{land}.}.
We present the result of
integration of the heavy modes in terms of 3d scalar and vector
fields with correct normalization of derivative terms in the
${\overline{MS}}$ scheme (in order to escape complicated notations we
keep the same notation for 4d and 3d fields):
\begin{eqnarray}
&&S_{\rmi{eff}}[A_i^a(\bfx),A_0^a(\bfx),\phi_i(\bfx)]
= \int d^3x \biggl\{{1\over4} F_{ij}^aF_{ij}^a +
\fr12 (D_iA_0)^a(D_iA_0)^a + (D_i\phi)^\dagger(D_i\phi)+
\nonumber
\\&&
\fr12 m_D^2 A_0^aA_0^a + \frac{1}{4}\lambda_A (A_0^aA_0^a)^2
+ m_3^2\phi^\dagger\phi
+\lambda_3(\phi^\dagger\phi)^2
+ h_3 A_0^aA_0^a \phi^\dagger\phi \biggr\}.
\label{3daction}
\end{eqnarray}
Here all bosonic fields have the canonical dimension [GeV]$^{1\over
2}$
and 3d gauge and scalar couplings  $g_3^2,~ \lambda_3,~\lambda_A$ and
$h_3$ have dimension [GeV]. These parameters and the masses $m_3$ and
$m_D$ can be expressed in terms of 4d couplings and the
temperature as follows:
\be
g_3^2 = g^2(\mu)T[1 + \frac{g^2 L_s}{(4 \pi)^2}(\frac{22}{3} -
\frac{1}{6}N_s)],
\label{g3}
\ee
\be
\lambda_3 = T[\lambda(\mu) - \frac{L_s}{(4 \pi)^2}(\frac{9}{16}g^4
-\frac{9}{2}\lambda g^2 + 12 \lambda^2)+\frac{1}{(4
\pi)^2}\frac{3}{8}g^4],
\label{lambda3}
\ee
\be
h_3 = \frac{1}{4}g^2(\mu)T[1 + \frac{g^2 L_s}{(4 \pi)^2}(\frac{22}{3}
-
\frac{1}{6}N_s) + \frac{1}{(4 \pi)^2}(12\lambda + \frac{49}{6}g^2
-\fr13 g^2N_s)]=
\label{h3}
\ee
\[\frac{1}{4}g_3^2[1 + \frac{1}{(4 \pi)^2}(12\lambda +
\frac{49}{6}g^2 -\fr13 g^2N_s)] ,
\]
\be
\lambda_A = \frac{17 g^4(\mu) T}{48 \pi^2},
\label{lambdaA}
\ee
\be
m_D^2 = \frac{5}{6} g^2(\mu) T^2,
\label{mD}
\ee
\be
m_3^2 = [{3\over16} g^2(\mu) + \fr12 \lambda(\mu)]T^2 - \fr12
m^2(\mu)[1 +
\frac{L_s}{(4 \pi)^2}(-\fr94 g^2 + 6 \lambda)],
\label{m3}
\ee
where
\be
L_s = 2\log{\mu e^\gamma\over4 \pi T}=\log{\mu^2\over T^2}-2c_B,\quad
c_B=\log(4\pi)-\gamma,
\ee
$\gamma$ is the Euler constant, $\mu$ is the scale parameter of the
${\overline{MS}}$ scheme and $N_s=1$ is the number of the scalar
doublets. Note the systematic appearance of the constant $c_B$ in
thermal
dimensional reduction.
We stress that all relations between coupling constants and
masses are gauge invariant (but may be scheme dependent).

The expressions of 3d fields in terms of bare 4d fields are  gauge
dependent, and we present them in Landau gauge. A number of terms can
be found in \cite{land}, we have added the contribution of scalars:
\be
\phi^{3d} = \frac{1}{\sqrt{T}}\phi(1-\frac{g^2}{(4
\pi)^2}\fr98\frac{1}{\epsilon_B}),
\ee
\be
A_0^{3d} = \frac{1}{\sqrt{T}}A_0[1+\frac{g^2}{(4 \pi)^2}
(\frac{5}{3} - \frac{13}{6}\frac{1}{\epsilon_B}+\frac{1}{12}
N_s(\frac{1}{\epsilon_B}+2))],
\ee
\be
A_i^{3d} = \frac{1}{\sqrt{T}}A_i[1+\frac{g^2}{(4 \pi)^2}
(- \frac{13}{6}\frac{1}{\epsilon_B}+\frac{1}{12}
N_s\frac{1}{\epsilon_B})].
\ee
Here
\be
1/\epsilon_B=1/\epsilon+L_s.
\label{epsB}
\ee

We note that the constants $g_3,~\lambda_3,~h_3$ as well as the term
proportional to $m^2$ in eq. (\ref{m3}) do not depend on the
parameter $\mu$ up to higher order terms in coupling constants.
The reason is that the coefficients in front of the logarithm $L_s$
are just the corresponding $\beta$-functions. This is not surprising,
as we will see in the next section. To this order of perturbation
theory there are no corrections proportional to $\log(\mu/T)$ to eqs.
(\ref{lambdaA},\ref{mD}) or to the term proportional to $T^2$ in eq.
(\ref{m3}). They can appear on the 2-loop level of dimensional
reduction, see Section 6. (We stress that no logs of the type
$\log(\mu/m)$ can appear when one integrates out heavy modes!) By the
choice
\be
\mu = \mu_T=4 \pi T e^{-\gamma}\sim 7 T
\label{muT}
\ee
all logarithmic
contributions can be removed. In loose terms, $\mu_T$ is an average
momentum of integration of heavy modes. Perturbation theory for the
{\em construction of the effective 3d theory from the 4d one} is
valid only when logarithmic corrections are small, i.e. with the
choice of parameter $\mu \sim \mu_T$. So, we have the explicitly
$\mu$
independent relations\footnote{Of course, the use of, e.g, $g(\mu_T)$
instead of $g(\mu)$ in the one-loop terms goes beyond the assumed
accuracy, but is not essential numerically.}
\[
g_3^2 = g^2(\mu_T)T,
\]
\[
\lambda_3 =
T[\lambda(\mu_T)+\frac{1}{(4\pi)^2}\frac{3}{8}g^4(\mu_T)],
\]
\[
h_3 =
\frac{1}{4}g_3^2[1 + \frac{1}{(4 \pi)^2}(12\lambda(\mu_T) +
\frac{49}{6}g^2(\mu_T) -\fr13 g^2(\mu_T)N_s)] ,
\]
\be
\lambda_A = \frac{17 g^4(\mu_T) T}{48 \pi^2},
\label{boundary}
\ee
\[
m_D^2 = \frac{5}{6} g^2(\mu_T) T^2,
\]
\[
m_3^2 = [{3\over16} g^2(\mu_T) + \fr12 \lambda(\mu_T)]T^2 - \fr12
m_H^2,~
m_H^2 \equiv m^2(\mu_T).
\]

Note that the coupling between $A_0$ and $\phi$ is not the same
as the coupling between $A_i$ and $\phi$.
The reason is the absence of the Lorents invariance at
non-zero temperatures. The difference, however, is very small, less
than $3$\% for Higgs masses $< 80$ GeV. In order to escape
unnecessary complications of the equations we use in what follows
\be
h_3 = \frac{1}{4}g_3^2
\label{h3new}
\ee
instead of (\ref{h3}).

Our aim is the construction of the effective potential at
sufficiently small $\phi$, so that particle masses coming from the
Higgs mechanism are much smaller than $\mu_T$. This should be done
{\em in effective 3d theory}, since the direct use of
perturbation theory and the ${\overline{MS}}$ scheme for the
effective potential at small $\phi$ in the underlying 4d theory will
immediately give rise to large logs. According to the philosophy of
the
effective theories \cite{weinberg} the equations (\ref{boundary}) are
to be used as boundary conditions for the determination of the
parameters of our 3d theory at small momenta (or, equally, fields
$\phi$).

The dimensional reduction of 4d gauge theory produces
non-renormalizable as well as renormalizable (in 3d) higher order
terms not included above. The relevant coupling constants are
suppressed by powers of temperature and by numerical constants. The
terms $\sim 1/T^3$ have been computed in \cite{jakovac}. For example,
the higher order term $\phi^\dagger\phi (A_0^aA_0^a)^2$ is multiplied
by $-\zeta(3)g_3^6/(64\pi^4T^3)$.
The coefficient of the renormalizable term $\sim (A_0^aA_0^a)^3$
vanishes
in SU(2) gauge theory but equals \cite{land} $\frac{889
\zeta(3)e^6}{6144 \pi^4}$ in quantum electrodynamics. It is clear
that the presence of these small terms cannot change the character of
infrared phenomena we are interested in.

In Sections 3--7 we will study in more detail the 3d effective
theory defined in (\ref{3daction}).

\section{Peculiarities of renormalization of 3d theory}
Until Section 5 we shall forget the 4d origin of our 3d theory
and study it as it stands as a 3d SU(2) gauge theory with a
fundamental
Higgs field $\phi$ and an adjoint Higgs field $A_0$.
On the tree level our theory
(\ref{3daction}) is characterized by the six parameters
$m_3,\,m_D,\,\lambda_A,\,\lambda_3,\,h_3$ and $g_3^2$. This 3d theory
is super-renormalizable. There are a finite number of irreducible
graphs which are ultraviolet divergent. One-loop diagrams are
presented in Fig. 1, and 2-loop diagrams in Figs. 2 ($\delta m^2$)
and 3 ($\delta m_D^2$). One-loop graphs are linearly divergent while
2-loop graphs are logarithmically divergent. All these diagrams
correspond to the mass renormalization of the triplet and doublet
Higgs fields. As usual in quantum field theory the appearance of
logarithmic divergencies introduces an extra scale, $\mu_3$. For
concreteness, we choose $\mu_3$ to be the scale parameter in the
modified minimal subtraction scheme ${\overline{MS}}$,
associated with dimensional
regularization. We stress that at this stage the parameter $\mu_3$
introduced here has nothing to do with the parameter $\mu$
introduced previously in the discussion of dimensional reduction.
The change in the parameter $\mu_3$ changes the physical content of
the theory unless tree parameters are varied with $\mu_3$ in some
consistent way, given by renormalization group equations. In our case
only mass diagrams contain logarithmic divergencies, so that the
coupling
constants $\lambda_A,~\lambda_3,~h_3$ and $g_3^2$ are 3d
renormalization group invariants. This is the reason for the
appearance of the $\beta$-functions in eqs.(\ref{g3}-\ref{m3}) of the
previous
section. Just on dimensional grounds one can write the following
renormalization group equations for the mass parameters:
\be
\mu_3 \frac{\partial m_3^2(\mu_3)}{\partial \mu_3}= -\frac{1}{16
\pi^2}f_{2m}, \quad
\mu_3 \frac{\partial m_D^2(\mu_3)}{\partial \mu_3}= -\frac{1}{16
\pi^2}f_{2D},
\ee
where $f_{2m}$ and $f_{2D}$ are second order polynomials in coupling
constants. We stress that these equations are exact due to the
super-renormalizable character of 3d gauge theory. Generally, to make
a theory finite, one has to add mass counterterms to the action
(\ref{3daction}),
\be
\delta S = \int d^3x \{ \delta m^2 \phi^\dagger\phi + \fr12 \delta
m_D^2 A_0^aA_0^a \},
\label{mctbeg}
\ee
where
\be
\delta m^2 = f_{1m} \Sigma + f_{2m} \Sigma_\rmi{log},~
\delta m_D^2 = f_{1D} \Sigma + f_{2D} \Sigma_\rmi{log},\label{efs}
\ee
where $f_{1m}$ and $f_{1D}$ are first order polynomials in coupling
constants. The quantity $\Sigma$ is related to the following linearly
divergent integral
\be
\int{d^3p\over (2\pi)^3}{1\over \bfp^2 + m^2}.
\ee
Clearly, it depends on the regularization scheme.
In the momentum regularization scheme ($\Lambda$ is an UV cutoff)
$\Sigma = \Lambda/2\pi^2$, in lattice regularization
$\Sigma = 0.252731/a$ with $a$ being the lattice spacing, and
$\Sigma = 0$ in the minimal subtraction scheme ${\overline{MS}}$.
The quantity $\Sigma_\rmi{log}$ is related to
the logarithmically divergent 2-loop sunset diagram ($d=3-2\epsilon$)
\begin{eqnarray}
H(m_1,m_2,m_3)&=&
\mu_3^{4\epsilon}\int {d^d p\over (2\pi)^d}{d^d k\over (2\pi)^d}
{1\over (\bfp^2+m_1^2)(\bfk^2+m_2^2)(|\bfp+\bfk|^2+m_3^2)}=\\
&=&{1\over64\pi^2\epsilon} + \frac{1}{(4 \pi)^2}(
\log\frac{\mu_3}{m_1 + m_2 + m_3} +\fr12).
\nonumber
\end{eqnarray}
Again, in momentum regularization
\be
\Sigma_{log} =
{1\over16\pi^2}(\log{\Lambda\over \mu_3}-\fr12),
\ee
in the minimal subtraction scheme
\be
\Sigma_{log} = {1\over64\pi^2\epsilon},
\ee
and on the lattice a numerical estimate of the sunset graph gives
\be
\Sigma_{log}= {1\over16\pi^2}\biggl(\log\frac{6}{a\mu_3} +
0.09\biggr).
\label{mctend}
\ee
The constant finite terms in these expressions are
defined in such a way that after renormalization the value of the
sunset graph is the same for all renormalization schemes we
discussed.

For a number of applications (in particular for the analysis of the
lattice data) the knowledge of the counterterms is a must. A direct
computation with the use of the relation (\ref{h3new}) gives:
\be
f_{1m} = -(\frac{9}{4} g_3^2 + 6\lambda_3),~f_{1D} =
-5(g_3^2+\lambda_A),\label{1loopcounter}
\ee
and for the logarithmic counterterms
\be
f_{2m} = {81\over16}g_3^4+9\lambda_3 g_3^2-12\lambda_3^2,~
f_{2D} = 5 \lambda_A^2 - 20 g_3^2 \lambda_A.
\label{count}
\ee
The expressions for $f_{1m},~f_{2m},~f_{1D}$ and $f_{2D}$ are {\em
exact} and do not have any higher order corrections. The derivation
of eqs.~(\ref{count}) is given in Appendix A. It is also quite
interesting to express $f_{2m}$ as a function of $m_H$ using the tree
relations $g_3^2=g^2T$, $\lambda_3=\lambda
T,~\lambda=g^2m_H^2/(8m_W^2)$:
\be
f_{2m} \propto \biggl[1-\biggl({m_H\over3m_W}\biggr)^2\biggr]
\biggl[3\biggl({m_H\over3m_W}\biggr)^2+1\biggr].
\label{f2m_mh}
\ee
It thus vanishes for $m_H=3m_W$!

To summarize, the theory under consideration is characterized by two
renormalization invariant scales $\Lambda_m$ and $\Lambda_D$, and
three independent coupling constants $g_3^2, \lambda_3$ and
$\lambda_A$ (since we used relation between $h_3$ and
$g_3$)\footnote{We stress that these relations are not spoiled by the
renormalization due to super-renormalizable character of 3d theory.}.
The relations between lagrangian masses and renormalization invariant
scales are given by
\be
m_3^2(\mu_3)=  {1\over16\pi^2}f_{2m}
\log\frac{\Lambda_m}{\mu_3},\quad
m_D^2(\mu_3) =  {1\over16\pi^2}f_{2D} \log\frac{\Lambda_D}{\mu_3},
\label{mLambdam}
\ee

Quite an interesting situation arises when $\lambda_A = 0$. Then
there is no 2-loop counter term for the $A_0$ mass term ($f_{2D}
=0$), the various diagrams in Fig. 2 cancel each other. Furthermore,
this cancellation also takes place in a 3d theory even without the
doublet
Higgs field. So, in this case $m_D^2$ is a renormalization
invariant quantity ($\mu_3$ independent). In fact the relation
$\lambda_A = 0$ is almost true for a three dimensional theory derived
by dimensional reduction from 4d high temperature one, since
according to (\ref{lambdaA}) $\lambda_A \sim 1.6 \cdot 10^{-2}
g_3^2$.
For a realistic choice of SU(2) gauge coupling $g = 2/3$
this is numerically much smaller than all other couplings. These can
as above
be estimated, say, from
$\lambda_3 = m_H^2/(8 m_W^2)g_3^2$. For QCD clearly $g$
and correspondingly $\lambda_A$ will be much larger and it is
possible that 2-loop counterterms could modify the studies of
dimensional reduction in hot QCD in \cite{reisz1}-\cite{reisz2}.

\section{The effective potential of the 3d theory to 2 loops}
The object which is used for the study of the phase transitions is
the effective potential \cite{jackiw}. As usual for gauge theories,
the
effective potential is gauge dependent. We use the Landau gauge and
dimensional regularization for all our computations.

In the 1-loop approximation the effective potential for the scalar
field in the effective 3d theory defined by the action
in eq.~(\ref{3daction})
was given in \cite{kari}. We define the mean field dependent
masses as
\begin{eqnarray}
&& m_T=\fr12 g_3\phi,\qquad\quad m_L^2= m_D^2 + \fr14 g_3^2\phi^2,
\nonumber \\
&& m_1^2=m_3^2(\mu_3)+3\lambda_3\phi^2,\quad
 m_2^2=m_3^2(\mu_3)+\lambda_3\phi^2
\la{masses}
\end{eqnarray}
Then the 1-loop potential is
\begin{eqnarray}
V_1(\phi) &=& \fr12 m_3^2(\mu_3)\phi^2
+\fr14\lambda_3\phi^4 -\nonumber \\
&& -{1\over12\pi} \biggl(6m_T^3+3m_L^3 + m_1^3+3m_2^3 \biggr).
\la{1looppot}
\end{eqnarray}
As one could expect, this potential coincides with the high
temperature expansion of potential in the
daisy wheel approximation found in a number of papers
\cite{carrington} -\cite{bhw}. The 1-loop computation of the
effective potential
cannot fix the arbitrary scale $\mu_3$ appearing as the argument in
the
scalar mass\footnote{The effective potential in the 1-loop
approximation is renormalization invariant only up to terms
of the order
of $g_3^4, g_3^2 \lambda_3$ and $\lambda_3^2$.}. For physical reasons
it is clear, though, that in order to minimize higher order
corrections this scale should be of the order of particle masses
defined by eq.~(\ref{masses}). We will return to the discussion of
this question later.

To fix the scale $\mu_3$ we must compute the 2-loop potential. The
relevant diagrams are given in Fig. 4 and the computation is
carried out in considerable detail in Appendix B.
Using the function
\be
\bar H(m_1,m_2,m_3)=\log{\mu_3\over m_1+m_2+m_3}+\fr12
\la{hbar}
\ee
the result for the 2-loop contribution to the potential in
dimensional
regularization is
\begin{eqnarray}
&&V_2(\phi,\mu_3)={1\over 16\pi^2}\biggl\{\nonumber\\
&&-{3g_3^4\over16}\phi^2 \biggl[
        2\bar H(m_1,m_T,m_T)-\fr12 \bar H(m_1,m_T,0)+
\bar H(m_1,m_L,m_L)\nonumber\\
&&\quad\qquad+{m_1^2\over m_T^2}[\bar H(m_1,m_T,0)-\bar
H(m_1,m_T,m_T)]\nonumber\\
    &&\quad\qquad +{m_1^4\over 4m_T^4}
[\bar H(m_1,0,0)+\bar H(m_1,m_T,m_T)-2\bar H(m_1,m_T,0)]\nonumber\\
&&\quad\qquad-{m_1\over 2m_T}-{m_1^2\over 4m_T^4}\biggr]\nonumber\\
&&-3\lambda_3^2\phi^2[\bar H(m_1,m_1,m_1)\bar
H(m_1,m_2,m_2)]\nonumber\\
&&+2g_3^2m_T^2[{63\over16}\bar H(m_T,m_T,m_T)+
{3\over16}\bar H(m_T,0,0)-{41\over16}]\la{2looppot}\\
&&\quad\qquad-\fr32 g_3^2[(m_T^2-4m_L^2)\bar
H(m_L,m_L,m_T)-2m_T m_L-m_L^2]\nonumber\\
&&+4g_3^2m_T^2+\fr38 g_3^2(2m_T+m_L)(m_1+3m_2) \nonumber\\
&&+ \frac{15}{4}\lambda_A m_L^2+\fr34\lambda_3(m_1^2+2m_1m_2+5m_2^2)
\nonumber \\
&&-\fr38 g_3^2\bigl[ (m_T^2-2m_1^2-2m_2^2)\bar H(m_1,m_2,m_T)+
(m_T^2-4m_2^2)\bar H(m_2,m_2,m_T)
\nonumber\\
&&\quad\qquad+{(m_1^2-m_2^2)^2\over m_T^2}[\bar H(m_1,m_2,m_T)-
\bar H(m_1,m_2,0)]\nonumber\\
&&\quad\qquad+(m_1^2-m_2^2)(m_1-m_2)/m_T+m_T(m_1+3m_2)-m_1m_2-m_2^2
\bigr]
\biggr\}. \nonumber
\end{eqnarray}

The logarithmically divergent 2-loop counter term in eq.(\ref{count})
can be directly extracted from eq.(\ref{2looppot}): the coefficient
$f_{2m}=81g_3^4/16+9\lambda_3 g^2 -12\lambda_3^2$ is simply the sum
of the coefficients of the $\fr12 \phi^2\bar H$ - terms within the
braces
there. The derivation of eq.~(\ref{2looppot}) is quite lengthy but it
is still much more straightforward than that in the 4d case at
non-zero
temperatures \cite{ae}-\cite{hebecker}\footnote{Shortly after our
paper was completed the paper by Z. Fodor and A. Hebecker (Preprint
DESY 94-025) appeared where the authors computed the high temperature
asymptotic of the two-loop effective potential accounting for
$\lambda^2$ and $g^2 \lambda$ terms. Due to different normalization
conditions we were not able to compare our results.}.

The total 2-loop effective potential, normalized to zero,
\be
V_\rmi{tot} = V_1(\phi) + V_2(\phi)- V_1(0) - V_2(0),
\label{vtot}
\ee
is now renormalization invariant up to the higher order terms in the
gauge and scalar self-coupling constants. Due to the fact that the
ultraviolet renormalization of any of the wave functions of the
fields in our 3d theory is absent, the exact effective potential,
normalized as in (\ref{vtot}), is renormalization group invariant.

In what follows, we will put for simplicity $\lambda_A = 0$, since
its numerical value is small. However, the more general case can be
treated also without any problems.

\section{Renormalization group and convergence of perturbation
theory}
There are two aspects of the convergence of perturbation theory
in 3d. The first problem is associated with ultraviolet logarithmic
renormalization. As far as we know it has not been discussed in the
literature in the present context. The second one is the well known
infrared problem. In spite of the fact that this problem has been
discussed already in a number of papers, we will discuss it too in
the Section 7, adding some new views and estimates.

\subsection{Preliminaries}
Since the 1-loop as well as the 2-loop effective potentials depend
explicitly on the scale $\mu_3$, while the exact effective potential
is $\mu_3$ independent, the question of the convergence of
perturbation theory must be formulated accounting for this fact. Let
us discuss this in more detail.

Clearly, this type of situation is not new. It arizes in any
calculation in any quantum field theory containing logarithmic
divergencies, in particular in QCD or QED. For example, the
cross-section of, say, gluon-gluon scattering in the tree
approximation is proportional to $\alpha_s^2(\mu)$ and {\em is not
defined} at all without the fixing of the scale $\mu$. The method of
dealing with this problem is, of course, well known. It is a {\em
combination} of the perturbation theory and {\em renormalization
group}, which sums up leading logarithmic corrections. In our example
with gluon scattering one simply says that $\mu^2 \sim Q^2$, where
$Q^2$ is a typical momentum transfer and uses a 1-loop running
coupling $\alpha_s$. Then, 1-loop corrections to the process do not
contain any logarithms of the type $\log(Q^2/\mu^2)$, so that the
expansion proceeds in powers of $\alpha_s(Q^2)$ rather than of the
quantity $\alpha_s(\mu)$. The question of the convergence of
perturbation theory becomes therefore more complicated. In
particular, for the processes depending on just one energy scale, the
parameter $\mu$ may be chosen in such a way that 1-loop
corrections to a tree result simply vanish. Suppose this happens at
some $\mu^2 = \kappa Q^2$. Then the QCD folklore says that
perturbation theory works if $\alpha_s(\kappa Q^2)/\pi \ll 1$. In
order to check the convergence of perturbation theory, one has to go
generally to a 2-loop level.

Of course, our 3d theory is different from theories in 4 dimensions.
Our gauge and scalar couplings do not run. The only quantity
depending on the scale $\mu_3$ is the mass of the scalar doublet. It
has a qualitatively different behaviour depending on the parameters
of the theory. As seen from eqs.~(\ref{f2m_mh}-\ref{mLambdam}), for
small scalar self-coupling (corresponding to $m_H < 3 m_W$)
$m_3^2(\mu_3)$ is positive at small $\mu_3$ and negative at large
$\mu_3$, while for $m_H > 3m_W$ the situation is opposite. The point
$m_H = 3 m_W$ is quite specific, for this value of the Higgs mass
logarithmic renormalization is absent. It is interesting to note that
in the abelian Higgs model
$f_{2m}=-6g_3^4[1-2m_H^2/(3m_W^2)+m_H^4/(3m_W^4)]$ is always negative
independently of the mass of the Higgs boson (see Appendix B.3).

\subsection{Example: The scalar theory}
In order to understand in more detail how the $\mu_3$ dependence
should
be dealt with, we take first a simplest possible example, namely
$\phi^4$ theory in 3d, defined by the Lagrangian
\be
{\cal L}= \fr12(\partial_i\phi)^2+\fr12m_3^2\phi^2+\fr14
\lambda_3\phi^4.
\ee
This theory is super-renormalizable and in the dimensional
regularization scheme there is only one divergent graph, namely the
sunset diagram contributing to the mass of the scalar field. The
2-loop effective potential, derived in detail in Appendix B.1, is:
\begin{eqnarray}
V_\rmi{eff}(\phi)&=& \fr12 m^2(\mu_3) \phi^2 + \fr14 \lambda_3\phi^4
-{1\over12\pi}m_1^3(\phi) +
\label{Vsc2loop}\\
&&+{\lambda_3\over64\pi^2}3m_1^2(\phi)
-3{\lambda_3^2\over16\pi^2}
(\log{\mu_3 \over3m_1(\phi)}+\fr12)\phi^2,\nonumber
\end{eqnarray}
where $m_1(\phi)^2=m^2(\mu_3)+3\lambda_3\phi^2$ and
\be
m^2(\mu_3) =  3{\lambda_3^2\over 8\pi^2} \log{\mu_3 \over \Lambda_m}.
\ee
The 3d theory depends on the parameters $\lambda_3$ and $\Lambda_m$.
The former is related to the 4d $\lambda$ via equation similar to
(\ref{lambda3}). We shall in Section 6 give $\Lambda_m$ in terms of
$T$.

Due to the $\mu_3$ dependence of $m(\mu_3)$, the 2-loop effective
potential is, in fact, $\mu_3$ independent up to 3-loop corrections.
The scale $\mu_3$ can thus be chosen at will and should be chosen
so as to minimize the effect of large logarithms.
For example, one may require that two-loop contribution to the
effective potential vanishes. In our case this happens at $\mu_3 =
3\exp(1/4)m_1(\phi)$. This simple prescription, however, is not
absolutely correct.
The reason is that the effective potential itself is not a good
object for
renormalization group studies. It is not a measurable physical
quantity -- only the difference
\be
V(\phi_1)-V(\phi_2)
\ee
has physical meaning. The potential is defined up to an additive
constant. In fact in the minimal subtraction scheme this constant
is $\mu_3$-dependent, so that the potential itself (not normalized to
zero at $\phi=0$) in  is not a
renormalization group invariant.
If $\phi_1$ and $\phi_2$ are very
different from each other, then characteristic momenta of integration
defining $V(\phi_1)$ and $V(\phi_2)$ are different, and
renormalization group improvement is not possible. The simplest way
to solve this problem is to consider the derivative of the effective
potential,
\be
DV(\phi) \equiv d V/d\phi
\ee
which must be $\mu_3$
independent. Moreover, the loop integration momenta are defined by a
single scale, $p \sim m(\phi)$.

The renormalization group equation is trivial:
\be
(\mu_3\frac{\partial}{\partial \mu_3}-\frac{1}{16 \pi^2}f_{2m}
\frac{\partial}{\partial
m^2})DV(\phi,\mu_3,m^2)=0.
\ee
The summation of the leading logs will be achieved with the choice of
\be
\mu_3 = 3 \kappa m_1(\phi),
\label{mu}
\ee
after first taking the derivative with respect to $\phi$,
where $\kappa$ is a constant determined by some criterion of
minimizing 2-loop effects. For example, one may require the
value of $\phi$ at the broken minimum at $T_c$ to be equal for
the 1- and 2-loop potentials. The renormalization group improved
expression for the effective potential is therefore
\be
V_\rmi{rg}(\phi) = \int_0^{\phi}d\phi DV(\phi,
\mu_3,m^2(\mu_3))|_{\mu_3=3\kappa m_1(\phi)}.
\label{improved}
\ee
with $\kappa$ being a number of order of one.

In fact, the knowledge of the effective potential up to two loops
allows one to compute 3- and 4-loop logarithmic terms with the help
of the renormalization group. Because of the $\mu_3$ dependence of
$m_1(\phi)$, there in $V_1$ is a $\mu_3$ dependence $\mu_3\partial
V_1/
\partial\mu_3$ which is of 3-loop order. For the 3-loop potential to
satisfy $\mu_3\partial V_3/\partial\mu_3=0$ it must contain a
compensating logarithmic term $-\mu_3\partial
V_1/\partial\mu_3\times\log[\mu_3/m_1(\phi)]$.
Thus the logarithmic term in the 3-loop correction to the potential
is
\be
\Delta V_3 = -\mu_3\frac{\partial
V_1(\phi)}{\partial\mu_3}\log{\mu_3\over
m_1(\phi)}={1\over12\pi}\mu_3{\partial m_1^3(\phi)\over\partial\mu_3}
\log{\mu_3\over m_1(\phi)}=
\frac{3\lambda_3^2}{(4 \pi)^3}m_1(\phi)\log{\mu_3 \over m_1(\phi)}.
\ee
Similarly, the 4-loop correction must contain the logarithmic term
\be
\Delta V_4 = -{\partial V_2\over\partial m^2}
\mu_3{\partial m^2\over\partial\mu_3}\log{\mu_3 \over m_1(\phi)}
=-\frac{3\lambda_3^3}{(4 \pi)^4} \biggl[\fr52 - {m^2
\over m_1^2(\phi)}\biggr]\log{\mu_3 \over m_1(\phi)}.
\ee
The comparison of,
say, 2-loop and 3-loop logarithmic terms gives an idea of when
renormalization group improved perturbation theory works, namely
for
\be
m_1(\phi)\gg \frac{\lambda_3}{4 \pi}
\ee
where $\mu_3$ is chosen in accordance with (\ref{mu}).

The summary of this exercise is as follows. Due to the
$\mu_3$-dependence
of the scalar mass tree and 1-loop potentials do not have definite
sense, since any result obtained with them contains unphysical
$\mu_3$
dependence. The scale $\mu_3$ can be fixed from the requirement that
the contribution of 2-loop diagrams is minimized. This happens when
$\mu_3$ is of the order of a typical scale in the problem, namely the
$\phi$ dependent scalar mass. In the 2-loop effective potential
$\mu_3$ dependence appears on the 3-loop level. The effective
summation of the leading logs can be achieved with a choice of
$\mu_3$
in accordance with eq.(\ref{mu}) on the level of 2-loop effective
potential.

\subsection{SU(2) theory}
The general remarks applicable to the pure scalar model are also
valid for SU(2). Namely, the 1-loop effective potential does not
have a definite sense without fixing of the scale $\mu_3$. The
dependence of
the 2-loop effective potential on $\mu_3$ appears on the 3-loop
level. However, the choice of $\mu_3$ is not so evident as in the
previous case. The reason is that we have now several different
scales in the problem, namely four different masses
($m_1$ and $m_2$ for Higgses, $m_T$
for the gauge boson and $m_L$ for the adjoint Higgs field).
For the applications to a realistic electroweak theory two cases are
the most interesting.

The first one is the case when the Higgs mass is so small
that the Higgs contributions to the effective potential
are suppressed by the small scalar self-coupling constants and
may be neglected. Then basically only $m_T$ and $m_L$ appear in the
logs (this happens also when the Higgs mass is of the order of $W$
mass, so that remarks below are applicable to this case as well).
The natural
guess is to take $\mu_3$ such that the sum of all 2-loop logarithmic
corrections vanishes. Neglecting scalar masses we get
\be
\mu_3 = \kappa m_L(\phi)^{1-\zeta} m_T(\phi)^{\zeta}\equiv \mu(\phi)
\label{musmall}
\ee
with
\be
\zeta=\frac{f_{2m}^0}{f_{2m}},
\ee
where $f_{2m}^0$ is to be computed in a theory without the triplet
Higgs field $A_0^a$ (see Section 5.4 and Appendix B.4):
\be
f_{2m}^0 ={51\over16}g_3^4+9\lambda_3 g_3^2-12\lambda_3^2,
\label{f2m0}
\ee
and $\kappa$ is some arbitrary parameter, $\kappa \sim 1$.
While this choice is certainly good for the minimization of the
2-loop corrections, one may wonder if it is good for the
higher order terms.
Later we will provide more motivation for this choice, coming from
the consideration of an effective theory with the $A_0^a$ field
integrated out.

So, we define a renormalization group improved effective potential in
a perfect analogy with (\ref{improved}),
\be
V_\rmi{rg}(\phi) = \int_0^{\phi}d\phi DV(\phi,
\mu_3,m^2(\mu_3))|_{\mu_3=\mu(\phi)}.
\label{improvedsu2}
\ee
where $DV$ is a partial derivative of the 1- or 2-loop effective
potential with respect to $\phi$.

The 2-loop effective potential depends on the scale $\mu_3$ only
weakly. The same is true for the dependence on $\kappa$ of the
renormalization group improved effective potential derived in the
2-loop
approximation. The 1-loop improved effective potential depends on
$\kappa$ strongly. The minimal sensitivity requirement would be to
choose $\kappa$ in such a way that the minimum of the 1-loop
effective potential is at the same point as for 2-loop potential.

The second case is when the Higgs mass is so large that $m_1$ and
$m_2$ appearing in the log terms are greater than the vector boson
mass.
Then the parameter $\mu_3$ should be associated with the scalar
masses.
We will not consider this case, since it is expected that the
electroweak phase transition is weakly first order in this region.

\subsection{Integrating out the $A_0$ field}
Going from 4 to 3 dimensions we were able to escape the problem of
large logs of the type $\log(T/m_L)$ and $\log(T/m_T)$. However, if
$m_T \ll m_L$, then logs of the type $\log(\mu_3/m_L)$ and
$\log(\mu_3/m_T)$ will appear on the level of 3d theory as well. The
effective theories provide again the recipe of the summation. One has
just integrate out the $A_0$ field and construct a new effective
theory (see Appendix B.4).

The new effective action in the 1-loop approximation is
\be
S_\rmi{eff}[A_i^a(\bfx),\phi_i(\bfx)]
= \int d^3x \biggl\{ {1\over4} F_{ij}^aF_{ij}^a +
 (D_i\phi)^\dagger(D_i\phi)+ \bar m_3^2\phi^\dagger\phi
+\bar \lambda_3(\phi^\dagger\phi)^2 \biggr\}.
\label{effective}
\ee
There is
no wave function renormalization coming from $A_0$ integration
and the connection between new and old parameters on the 1-loop level
is
\be
\bar g_3^2 = g_3^2(1 - \frac{g_3^2}{24 \pi m_D}),
\label{g3eff}
\ee
\be
\bar \lambda_3 = \lambda_3 - \frac{3 g_3^4}{128 \pi m_D},
\label{lameff}
\ee
\be
\bar m_3(\mu_3)^2 = m_3^2(\mu_3) - \frac{3 g_3^2 m_D}{16\pi}
+\frac{15 g_3^4}{8(4\pi)^2}(\log\frac{\mu_3}{2 m_D} +\frac{3}{10}).
\label{meff}
\ee
The last term in the $\bar m_3^2(\mu_3)$ removes the heavy $A_0$
degrees of freedom from the
renormalization group equation, i.e., replaces $f_{2m}$ by $f_{2m}^0$
defined in eq.~(\ref{f2m0}). We shall derive the above relations in
Appendix B.4 by considering the $m_D\to\infty$ limit of the 2-loop
potential. The 2-loop effective potential for
the theory defined by (\ref{effective}) can be
extracted from (\ref{2looppot}) just by dropping the $A_0$
contributions and relating the parameters by
eqs.(\ref{g3eff}-\ref{meff}).
We present the result for completeness:
\begin{eqnarray}
&&V(\phi)= \fr12 \bar m_3^2(\mu_3)\phi^2
+\fr14\bar \lambda(\phi^2)^2 \nonumber \\
&& -{1\over12\pi} \bigl(6m_T^3+ m_1^3+3m_2^3 \bigr)
+{1\over16\pi^2}\biggl\{
\nonumber\\
&&-{3\bar g_3^4\over16}\phi^2 \biggl[
        2\bar H(m_1,m_T,m_T)-\fr12 \bar H(m_1,m_T,0)\nonumber\\
&&\quad\qquad+{m_1^2\over m_T^2}[\bar H(m_1,m_T,0)-\bar
H(m_1,m_T,m_T)]\nonumber\\
    &&\quad\qquad +{m_1^4\over 4m_T^4}
[\bar H(m_1,0,0)+\bar H(m_1,m_T,m_T)-2\bar H(m_1,m_T,0)]\nonumber\\
&&\quad\qquad-{m_1\over 2m_T}-{m_1^2\over
4m_T^4}\biggr]\nonumber\\
&&-3\bar \lambda_3^2\phi^2[\bar H(m_1,m_1,m_1)+\bar
H(m_1,m_2,m_2)]\nonumber\\
&&+2\bar g_3^2m_T^2[{63\over16}\bar H(m_T,m_T,m_T)+
{3\over16}\bar H(m_T,0,0)-{41\over16}]-\la{va0}\\
&&+4\bar g_3^2m_T^2+\fr34 \bar g_3^2m_T(m_1+3m_2)
+\fr34\bar \lambda_3(m_1^2+2m_1m_2+5m_2^2)
\nonumber \\
&&-\fr38 \bar g_3^2\bigl[ (m_T^2-2m_1^2-2m_2^2)\bar H(m_1,m_2,m_T)+
(m_T^2-4m_2^2)\bar H(m_2,m_2,m_T)
\nonumber\\
&&\quad\qquad+{(m_1^2-m_2^2)^2\over m_T^2}[\bar H(m_1,m_2,m_T)-
\bar H(m_1,m_2,0)]\nonumber\\
&&\quad\qquad+(m_1^2-m_2^2)(m_1-m_2)/m_T+m_T(m_1+3m_2)-m_1m_2-m_2^2
\bigr]
\biggr\}, \nonumber
\end{eqnarray}
The mass parameters appearing in this expression are to be computed
with the help of eq. (\ref{masses}) and substitution $g_3 \rightarrow
\bar g_3$,
$\lambda_3 \rightarrow \bar \lambda_3$, $m_3 \rightarrow \bar m_3$.
To get a renormalization group improved potential one now just
chooses $\mu_3 \sim m_T$ in the effective theory without the $A_0$
field and uses (\ref{improvedsu2}). Note that due to
(\ref{meff}) this choice of $\mu_3$ corresponds to (\ref{musmall}).
Again, no large logs will appear.
We note also that this procedure sums up in all orders of
perturbation theory $A_0$ ring insertions.

Now we are ready to discuss the relation between high temperature 4d
and 3d physics on the 2-loop level.

\section{The relation between 3d and 4d computations on the 2-loop
level}
In Section 2 we have derived the parameters of the effective 3d
theory in terms of the 4d parameters by explicit integration over
the nonstatic modes of the 4d theory. However, the calculation is not
yet complete: due to 3d renormalization the mass parameters are of
the
form
\be
m_3^2(\mu_3)=  {1\over16\pi^2}f_{2m}
\log\frac{\Lambda_m}{\mu_3},\label{m3general}
\ee
and we still have to express the scale $\Lambda_m$ in terms of $T$.
This can be done by comparing the computed 2-loop potentials in 3d
and in 4d \cite{ae}.

To make this comparison simple two facts can be taken into account.
First, we write the 4d analogue of the H-function as\footnote{The 4d
sunset diagram can be decomposed as the sum of 3 pieces, $H^{4d} =
H_{lll} + H_{hhh}+H_{hhl}$, where $H_{hhh}$ is contribution
containing only heavy 3d modes,  $H_{hhl}$ is a sum containing one
light 3d mode and 2 heavy, while $H^{lll}$ is 3d contribution defined
in (\ref{hbar}).
In ref.\cite{ae} it was incorrectly asserted that $H_{hhl}$ is
suppressed by the light mass. The direct computation of $H_{hhl}$ in
the massless limit gives
\be
H_{hhl}= -\frac{3}{64 \pi^2}[\frac{1}{\epsilon}+ 2(1 -\log 4 +
\log\frac{\mu^2}{T^2})].
\ee
This mistake, as far as we can see, does not change the
final results of ref.\cite{ae}.} (sf. eq. (3.17) of ref.\cite{ae})
\be
H^{4d}(m_1,m_2,m_3)={T^2\over
(4\pi)^2}({1\over4\epsilon_B}+\frac{1}{4}l_{\epsilon} +
\log{3T\over\mu_3}
+\bar{H}(m_1,m_2,m_3)+c),
\ee
where $\epsilon_B$ is given by (\ref{epsB}), the constant
$l_{\epsilon}$ is defined in eq. (3.13) of \cite{ae}(it cancels out
in the final result).
The numerical constant $c$ is related to $c_H$ computed numerically
in \cite{parwani}\footnote{The constant $c$ appears to
be amazingly close to $-\fr12 \log2$. However, we did not try to
prove that this is indeed the case.},
\be
c\equiv\fr14(2c_B - c_H)=-0.348725,\quad c_H=5.3025,\quad
c_B=\log(4\pi)-\gamma_E=1.9538.
\ee

Second, we notice that if we choose the 4d scale parameter $\mu$ as
in (\ref{muT}) then the constant $c_B$ does not appear in any of the
expressions defining high temperature asymptotic of the effective
potential.

With this in mind consider first the simplest case, O($N$) scalar
theory with $N$=1 discussed in the Section 5.2 with 4d lagrangian
\be
{\cal L}= \fr12(\partial_i\phi)^2+\fr12m^2\phi^2+\fr14
\lambda\phi^4.
\ee
Comparing first the coefficients of the $\phi^4$ terms in
eq.(\ref{Vsc2loop}) and in
eq.(3.29) of \cite{ae} gives the result
\be
\lambda_3=[\lambda(\mu)+{9\over16\pi^2}L_s\lambda^2]T
=\lambda(\mu_T)T
\ee
in analogy with eq.(\ref{lambda3}) with $\mu_T$ defined in
(\ref{muT}).
Note that the last term arises from the term $-2c_B$ in $L_s$. To
find out $m_3^2$ we compare the coefficients of the $\phi^2$ term
and express $\lambda(\mu_T)$ in terms of $\lambda_3$. This gives the
result
\be
m_3^2(\mu_3)=\gamma(T^2-T_0^2)+{1\over16\pi^2}\biggl[-6\lambda_3^2
\bigl(\log{3T\over\mu_3}+c \bigr)\biggr],
\label{m(mu)3d4d}
\ee
where
\be
\gamma(T^2-T_0^2)=\fr14\lambda(\mu_T)T^2-m^2(\mu_T).
\ee
Here the parameters $\lambda(\mu)$ and $m(\mu)$ are 4d quantities
determined in the 4d theory by the coupling strength and the Higgs
mass and by a choice of scale. In agreement with general relations,
the
coefficient of $\log(3T/\mu_3)$ is $f_{2m}/16\pi^2$ and the value of
$\Lambda_m$ in eq.(\ref{m3general}) can be read from
eq.(\ref{m(mu)3d4d}).
Note that the
only $\mu_3$-dependent 3d quantity is the effective mass of the
scalar
particle, the 4d $\mu$-dependence in the scalar self-coupling is
cancelled out.

According to eq.~(\ref{m(mu)3d4d}) the choice of scale
\be
\mu_3=3\exp(c)T= 2.117 T
\ee
makes the relation of $m_3^2(\mu_3)$ to 4d parameters be most simple.

The corresponding relations in the SU(2) case are more complicated.
Comparing with \cite{ae} we can now list the additional 4d
contributions which have to be added to our 3d result for the various
diagrams listed in Appendix
B:
\begin{eqnarray}
(d1)&=& -\frac{7}{16}g_3^2m_T^2,\nonumber\\
(d2)&=&\frac{39}{16}g_3^2m_T^2,\nonumber\\
(f2)&=&\frac{1}{16} g_3^2(4m_T^2+m_1^2+3m_2^2),\nonumber\\
\end{eqnarray}
All other diagrams give the same contribution in 3d and in high
temperature limit of 4d.

There are also two counter term diagrams (wave function
renormalization of the vector field and renormalization of the
$\phi^{\dagger}\phi (A_i^a)^2$ vertex) without a 3d counterpart
contributing $59g_3^2m_T^2/48$.

With this one can get \footnote{In order to escape confusion, we note
that in ref. \cite{ae} the result for the effective potential is
given in terms of 4d field normalized at scale $T$. This field is
related to our 3d field as
\be
\phi^2(T)=T\phi_3^2(1-{9g^2\over32\pi^2}c_B).
\ee}
\[
m_3^2(\mu_3) = [{3\over16} g^2(\mu_T) + \fr12 \lambda(\mu_T) +
\frac{g^2}{(4 \pi)^2}(\frac{167}{96}g^2 + \fr34 \lambda)]T^2
\]
\be
- \fr12 m^2(\mu_T)+
{1\over16\pi^2}\biggl[
f_{2m}\bigl(\log{3T\over\mu_3}+c\bigr)\biggr]=
\label{m32loop}
\ee
\[
[{3\over16} g_3^2 T + \fr12 \lambda_3 T + \frac{g_3^2}{(4
\pi)^2}(\frac{149}{96}g_3^2 + \fr34 \lambda_3)] - \fr12
m_H^2+
{1\over16\pi^2}\biggl[
f_{2m}\bigl(\log{3T\over\mu_3}+c\bigr)\biggr].
\]
This relation is a central result of this Section.

\section{Parameters of the EW phase transition from
perturbation theory}
\subsection{The 2-loop results}
Now, with effective field theories and the renormalization group at
hand we are ready to present perturbative predictions for the
critical
temperature for the different Higgs masses. In our numerical
estimates we used:
\be
g_3^2 = (\frac{2}{3})^2 T,~\lambda_3 =
\frac{1}{8}g_3^2\frac{m_H^2}{m_W^2},~
h_3 = \frac{1}{4}g_3^2, ~\lambda_A = 0.
\ee
In other words, we have neglected 4d running of the couplings and
masses. The parameter $m_H$ is not the physical mass of the Higgs
boson
(defined by the pole of the scalar propagator) but the Lagrangian
parameter in the $\overline{MS}$ scheme at the normalization scale
$\mu_T$.
 As discussed in Section 2, the difference is quite small
numerically.

The results are summarized in Table 1. The first 2 columns of the
table provide the critical temperature and the expectation value
of the Higgs field derived from the potential (\ref{improvedsu2}), in
which $m_3^2(\mu)$ from eq.(\ref{m32loop}) is used.
The next column is the value of
the parameter $\kappa$ for which the expectation value
of the Higgs field at $T=T_c$ is
the same for the 1-loop and 2-loop potentials. Pictures of typical
potentials are presented in Fig.5 ($m_H = 35$ GeV) and Fig. 6 ($m_H =
80$ GeV). We stress that the deviation
of the 1-loop RG improved potential from the 2-loop
can be removed by change of the parameter $\mu$.

\begin{table}
\center
\begin{tabular}{|c|c|c|c|}
\hline
 $m_H$ & $T_c$ & $\phi_\rmi{c}$ & $\kappa$  \\
\hline
 $35$ GeV & 93.2 & 160.7 & 2.03  \\
\hline
 $60$ GeV & 140.3 & 96.6 & 1.91  \\
\hline
 $70$ GeV & 157.5 & 87.3 & 1.83  \\
\hline
 $80$ GeV & 173.3 & 81.0 & 1.75  \\
\hline
 $90$ GeV & 187.9 & 76.1 & 1.66  \\
\hline
\end{tabular}
\caption[1]{Numerical values of $T_c$ and $\phi_c$ for various $m_H$
values.
The last column gives the value of $\kappa$ in eq.~(\ref{musmall})
for which the 1- and 2-loop results coincide.}
\end{table}

In Table 2 we present the results for the theory where the heavy
$A_0$ component has been integrated out. One can see that the
difference is within $1$ \% for the critical temperature and
$\phi_c$. This proves that the $A_0$ field has small influence on the
electroweak phase transition and indicates that higher order
corrections associated with $A_0$ are numerically small.

\begin{table}
\center
\begin{tabular}{|c|c|c|c|}
\hline
 $m_H$ & $T_c$ & $\phi_{c}$ &$\kappa$  \\
\hline
 $35$ GeV & 93.3 & 164.9 & 3.1  \\
\hline
 $60$ GeV & 140.1 & 96.6 & 2.66  \\
\hline
 $70$ GeV & 157.0 & 88.0 & 2.50  \\
\hline
 $80$ GeV & 172.6 & 82.0 & 2.37  \\
\hline
 $90$ GeV & 186.9 & 77.7 & 2.21  \\
\hline
\end{tabular}
\caption[1]{As Table 1 but without the $A_0$ field.}
\end{table}

\subsection{Perturbative uncertainties in predictions}
The crucial question with perturbative computations is of course
whether
they are reliable or not. As we discussed above, perturbation
theory does not work at all at small scalar fields due to infrared
divergences and a reliable estimate of the critical temperature is
not possible. The effective potential has a logarithmic singularity
at the origin at the 4-loop level and has the power divergent
corrections at still higher loop levels. Just on dimensional
grounds one can write the following expression for the effective
potential coming from the loop expansion near the origin, starting
from four loops:
\be
V_{\geq 4}(\phi)= \frac{1}{2 \pi}(\frac{g_3^2}{2 \pi})^3\left(C_0
\log\frac{\mu_3}{m_T(\phi)} + \sum_{n=1}^{\infty}C_n \rho^n \right).
\label{4loop}
\ee
Due to the dimensionality of the gauge coupling constant the
expansion parameter $\rho$ is expected to be
\be
\rho = \frac{1}{2\pi}\frac{g_3^2}{m_T(\phi)},
\ee
where $2 \pi$ comes from the loop integration, $m_T(\phi)=\fr12
g_3\phi$
and $C_n$ are some numerical coefficients. This parameter is also the
ratio of the typical terms of the 2- and 1-loop potentials
calculated earlier.
Perturbation theory explodes at  $\phi = 0$ and is not applicable
at $\rho > 1$. A natural way to estimate the uncertainties
associated with the divergent behaviour of the perturbative
expression
for the effective potential near the origin is to {\em assume} that
higher order corrections are not essential at $\rho <1$ (i.e. at
$m_T(\phi)> g_3^2/2\pi$) {\em and to assume} that they cannot change
the value of the potential at the origin by more than $A/(2
\pi)(g_3^2/2 \pi)^3$ with $A \sim 1$ (cf. eq.(\ref{4loop})).
Of course, these assumptions are completely ad hoc and cannot be
proven by any perturbative methods. Moreover, as we will discuss in
our paper \cite{fkrs2}, they are most probably wrong. Nevertheless,
since our aim in this section is to estimate the {\em perturbative}
uncertainties, we will make these assumptions. In
any case, perturbative uncertainties are present and it is worthwhile
to
understand whether they are important numerically.

The explicit computation of the effective potential up to two loops,
however, cannot help to estimate the convergence of perturbation
theory. Quite an
interesting situation arises. We cannot decide whether perturbation
theory works well or not by comparing 1-loop and tree potentials,
since just
by chance the scalar self-interaction is smaller than the gauge one.
At
the same time, the comparison of the 1-loop result with the
2-loop one does not provide any information either, due to the
$\mu$-dependence (the 1-loop result is exactly equal to the
2-loop result with some particular choice of $\mu$!).

This leads to the conclusion that most important uncertainties should
come from the 3-loop corrections, since their contributions to the
effective potential are parametrically larger than 4- and higher loop
contributions.  Unfortunately, the 3-loop computation is absent at
present\footnote{The number of different 3-loop graphs is about
100. We thank S. Larin for a discussion of this point.}, so
that we can only guess what their magnitude can be. In what follows
we consider two different types of corrections.

\subsubsection{The renormalization group corrections}
First of all, we know the $\mu$-dependent logarithmic
contribution on the 3-loop and
even on the 4-loop level from the renormalization group.
The 3-loop logarithmic correction to the
effective potential is:
\be
\Delta V_3 = {1\over16\pi^2}\frac{\partial V_1(\phi)}{\partial
m_3^2}f_{2m}
\log({\mu_3 \over m_T(\phi)}) =
\ee
\[
-2({1\over 4\pi})^3(m_1+3m_2)({81\over16}g_3^4+9\lambda_3
g_3^2-12\lambda_3^2)\log({\mu_3 \over m_T(\phi)}),
\]
and the 4-loop logarithmic terms are
\be
\Delta V_4 = {1\over16\pi^2}\frac{\partial V_2(\phi)}{\partial
m_3^2}f_{2m}
\log({\mu_3 \over m_T(\phi)}),
\ee
where
\begin{eqnarray}
&&16 \pi^2 \frac{\partial V_2(\phi)}{\partial m_3^2}=
{{-9\,{g_3^2}}\over {16}} + {{9\,\lambda_3}\over {128}} +
  {{3\,{g_3^2}\,{{{{} m_1}}^2}}\over
    {8\,\left( 4\,{{{{} m_T}}^2} + 6\,{{} m_T}\,{{} m_1} +
2\,{{{{} m_1}}^2} \right) }} +
  {{3\,{g_3^3}\,\phi}\over {32\,{{} m_1}}} +\nonumber\\
&&
  {{3\,{g_3^3}\,{{} m_1}\,\phi}\over
    {8\,\left( 4\,{{{{} m_T}}^2} + 6\,{{} m_T}\,{{} m_1} +
        2\,{{{{} m_1}}^2} \right) }} +
  {{3\,\lambda_3\,{\phi^2}}\over {8192\,{{{{} m_1}}^2}}} +
  {{3\,{g_3^4}\,{\phi^2}}\over
    {16\,{{} m_1}\,\left( 2\,{{} m_T} + {{} m_1} \right) }} +
\nonumber\\&&
  {{3\,{g_3^4}\,{\phi^2}}\over
    {32\,{{} m_1}\,\left( 2\,{{} m_L} + {{} m_1} \right) }} -
  {{3\,{g_3^4}\,{\phi^2}}\over
    {32\,{{} m_1}\,\left( 2\,{{} m_T} + 2\,{{} m_1} \right) }}
+\nonumber\\&&
  {{3\,\lambda_3\,{\phi^2}}\over
    {8192\,{{} m_1}\,\left( {{} m_1} + 2\,{{} m_2} \right) }} +
  {{3\,\lambda_3\,{\phi^2}}\over
    {4096\,{{} m_2}\,\left( {{} m_1} + 2\,{{} m_2} \right) }}
-\nonumber\\&&
  {{3\,{{{{} m_1}}^2}\,\log ({{3\,\mu_3}\over {{{} m_1}}})}\over
    {2\,{\phi^2}}} - {{3\,{g_3^2}\,
      \log ({{3\,\mu_3}\over {{{} m_T} + {{} m_1}}})}\over 4} +
\nonumber\\&&
  {{3\,{{{{} m_1}}^2}\,\log ({{3\,\mu_3}\over {{{} m_T} + {{}
m_1}}})}\over
    {{\phi^2}}} + {{3\,{g_3^2}\,\log ({{3\,\mu_3}\over
         {2\,{{} m_T} + {{} m_1}}})}\over 4} -
\nonumber\\&&
  {{3\,{{{{} m_1}}^2}\,\log ({{3\,\mu_3}\over
         {2\,{{} m_T} + {{} m_1}}})}\over {2\,{\phi^2}}} +
  {{3\,{g_3^2}\,\log ({{3\,\mu_3}\over {{{} m_T} + {{} m_1} + {{}
m_2}}})}\over
     2} + {{3\,{g_3^2}\,\log ({{6\,\mu_3}\over
         {2\,{{} m_T} + 4\,{{} m_2}}})}\over 2}.
\end{eqnarray}
In spite of the fact that in some terms the square of the Higgs field
appears in the denominator, this expression is perfectly regular for
$\phi \rightarrow 0$. The calculation of still higher order
logarithmic terms is not possible with the knowledge of the 1- and
2-loop expressions. Now, these expressions can be used for the
estimation of a part of 3-loop corrections with a `sensible' choice
of $\mu_3 \sim m_T(\phi)$. In practice we just varied the parameter
$\kappa$ in eq.(\ref{improvedsu2}) from $0.6$ to $6$ (this is roughly
factor $3^{(\pm 1)}$ deviation from the "optimal" value defined in
Table 1. The
results of these variations are shown in Table 3. One can see that
the corrections are quite small.

\begin{table}
\center
\begin{tabular}{|c|c|c|c|c|}
\hline
 $M_H$ & $ T_c$,  $\kappa=0.6$  & $\phi_{c}$,
$\kappa=0.6$  & $ T_c$,   $\kappa=6$  &
$\phi_{c}$, $\kappa=6$   \\
\hline
 $35$ GeV & 93.2 & 160.6 & 93.1 & 160.7  \\
\hline
 $60$ GeV & 140.6 & 96.3 & 139.9 & 97.2  \\
\hline
 $70$ GeV & 157.9 & 86.9 & 156.8 & 88.5  \\
\hline
 $80$ GeV & 174.0 & 80.4 & 172.3 & 82.6  \\
\hline
 $90$ GeV & 188.7 & 75.3 & 186.4 & 78.6  \\
\hline
\end{tabular}
\caption[1]{Estimated effect of the 3-loop logarithmic correction to
the
perturbative values of $T_c$ and $\phi_c$. Here $\kappa_0$ is the
value of $\kappa$ given in Table 1.}
\end{table}
\subsubsection{The 3-loop linear term}
The logarithmic corrections we have discussed are associated with the
2-loop renormalization of the Higgs mass. There will be a plenty of
other contributions (and, most
probably, more important numerically), related to the
self-interaction
of the gauge particles. Just on dimensional grounds the 3-loop
contribution is linear in the Higgs field and may be parametrized as
\be
\Delta V_3 = \frac{\beta}{(4 \pi)^3}g_3^4 m_T(\phi),
\label{3loop}
\ee
with $\beta$ being some unknown number. To get an idea of how large
it
can be
let us take the theory without the $A_0$ field (as we saw, the $A_0$
influence is quite small) and compare the 1-loop gauge contribution
$v_1 \sim \frac{1}{2\pi}m_T^3$  with the leading 2-loop contribution
$v_2 \sim \frac{1}{(4\pi)^2}\frac{51}{8}g_3^2 m_T^2(\phi)$, assuming
that the log is of the order of 1. The expansion parameter $\rho$
from here is $\rho \sim v_2/v_1 \sim g_3^2/(\pi
m_T(\phi))$. So, one can guess that $v_3 \sim v_2 \rho \sim
v_2^2/v_1$, i.e. $\beta$ can be as large as $51^2/(8\cdot 16) \sim
20$. We understand, of course, that estimates of this type
can never be trusted, but we cannot do anything better.
A number of constraints on higher order terms can be derived from
the analysis of lattice data \cite{fkrs2}. We just note
here that negative values of $\beta$ are certainly excluded
\cite{fkrs2}.

In table 4 we present the results of a computation of $T_c$
and $\phi_c$ with the
effective potential to which the linear term (\ref{3loop}) is added.
We vary the constant $\beta$ from $0$ to $20$, positive values of
$\beta$ decrease the critical temperature and increase the
expectation value, while the effect of a negative $\beta$ is the
opposite. As one can see, the uncertainty induced by this correction
is considerably larger than from log type contributions but still
within $20$ \% for the vacuum expectation value of the Higgs field at
the critical temperature\footnote{For negative $\beta \sim -20$ the
correction is $100$ \% important.}.
\begin{table}
\center
\begin{tabular}{|c|c|c|}
\hline
 $M_H$ & $ T_c$  with $\beta=20$ & $\phi_{c}$   with $\beta=20$ \\
\hline
 $35$ GeV & 92.2 & 166.2   \\
\hline
 $60$ GeV & 137.5 & 111.0  \\
\hline
 $70$ GeV & 153.9 & 104.5  \\
\hline
 $80$ GeV & 168.9 & 100.3 \\
\hline
 $90$ GeV & 182.8 & 97.5 \\
\hline
\end{tabular}
\caption{Estimated effect of a 3-loop linear term.}
\end{table}
In order to avoid confusion, we also stress here that
the upper bound on the Higgs mass, following from the requirement
\cite{s:higgs}
$v/T >1.5$ {\em cannot} be extracted from the perturbative effective
potential at the perturbative critical temperature (as was done, e.g.
in \cite{bd,ae,ay})  since  the bubble nucleation temperature can be
substantially lower than $T_c$ \cite{ms}.

\section{Conclusions}
We have considered the effective field theory approach to the study
of
the finite $T$ electroweak phase transition. We have constructed
explicitly an effective 3d SU(2) gauge field + fundamental Higgs
+ adjoint Higgs effective theory and studied its renormalization in
detail.
The main result is
that the large logarithms found in refs. \cite{bd,ae} on the 2-loop
level
are in fact harmless and can be summed up in all orders of
perturbation
theory by standard renormalization group  methods applied to
the 3d effective theory.
We have further derived an effective theory by integrating
away the heavy adjoint $A_0$ field .

Our results indicate that the 3d effective field theory approach is
a very useful one for the perturbative study of hot electroweak
matter
in the broken phase. They form a necessary background for
non-perturbative
lattice Monte Carlo simulations of the system, which are required
to clarify the properties of the phase transition and the symmetric
phase.

\appendix
\section{Appendix A}

In this appendix we carry out the computation of the logarithmically
divergent 2-loop counter term. The result could be obtained as a
limit
of the 2-loop potential, but a separate calculation is both
illustrative and
useful as a check.

The relevant Feynman diagrams were shown in Figs.~1-3. Landau gauge
is used
for $A_i^a$ so that diagrams with a zero-momentum external scalar
line
emitting a single $A_i^a$ line vanish, only the 4-vertex remains.
For generality we give the results for SU($N_c$) gauge field coupled
to a
fundamental and adjoint Higgs. The relevant fundamental-adjoint
couplings
are then ($A_0=A_0^aT^a,\tr T^aT^b=\fr12\delta_{ab},
(T^aT^a)_{ij}=C_F\delta_{ij}$)
\be
g_3^2 \phi^\dagger A_0A_0\phi,\quad g_3^2 \phi^\dagger A_iA_i\phi ,
\ee
the adjoint-adjoint coupling is ($A_0^a$ is an adjoint vector,
$A_i=A_i^aF^a,
\tr F^aF^b=N\delta_{ab}$)
\be
\fr12 g_3^2 A_0A_iA_iA_0,
\ee
and the scalar self-couplings are
\be
\fr14\lambda_3 (\phi_k\phi_k)^2,\quad \fr14\lambda_A (A_0^aA_0^a)^2.
\ee

Consider first the 1-loop diagrams in Fig.~1. In the calculation of
the
2-loop diagrams we shall actually need the 1-loop diagrams
with an exact propagator $D(k)=1/[k^2+\Pi(k)]$,
$D_{ij}(k)=(\delta_{ij}-k_ik_j/k^2)/[k^2+\Pi_T(k)]$ in the loop.
Using the
above couplings to sum over all the contributing intermediate states
one obtains the following contributions to $\delta m^2$ from the
diagrams in Fig.~1:
\begin{eqnarray}
\delta m^2_{(a)}& =& 2 g_3^2C_F\Sigma(\Pi_T(k)),\nonumber\\
\delta m^2_{(b)} &=&  g_3^2C_F\Sigma(\Pi_0(k)), \nonumber\\
\delta m^2_{(c)}& =& 2(N_c+1)\lambda_3\Sigma(\Pi_\phi(k)),
\nonumber\\
\delta m^2_{D(d)} &=& 2 g_3^2N_c\Sigma(\Pi_T(k)), \\
\delta m^2_{D(e)} &=& (N_c^2+1)\lambda_A\Sigma(\Pi_0(k)), \nonumber\\
\delta m^2_{D(f)} &=&  g_3^2\Sigma(\Pi_\phi(k)),\nonumber
\end{eqnarray}
where we have abbreviated
\be
\Sigma(\Pi(k))=\int {d^3k\over (2\pi)^3}{1\over k^2+\Pi(k)}.
\ee
For the 1-loop diagrams $\Pi(k)=0$  and the counter terms in
eq.(\ref{1loopcounter}) are obtained after taking $N_c=2, C_F=3/4$.

The 2-loop diagrams in Figs.~2-3 are of two different types, the
sunset
graphs (a)-(c) and the improved 1-loop graphs (d)-(h). Consider first
their general structure neglecting the exchange and internal symmetry
factors.

The sunset graphs come in two different variants, one with three
scalars in the intermediate state, the other with two vectors and one
scalar. The former is obviously proportional to
\be
H_\rmi{SSS}\equiv H ={1\over16\pi^2}\frac{1}{\epsilon},
\ee
while for the latter (in Landau gauge) we find
\be
H_\rmi{SVV}= \int {d^3p\over(2\pi)^3}{d^3q\over(2\pi)^3}
{1\over \bfp^2\bfq^2|\bfp+\bfq|^2}\biggl[1+{(\bfp\cdot\bfq)^2 \over
\bfp^2\bfq^2} \biggr] = \fr32 H.
\ee

The improved 1-loop graphs lead to a logarithmic divergence due to
the fact that (already for for dimensional reasons) the 1-loop
self energy $\Pi(k)$
depends linearly on $k$ in 3d: $\Pi(k)=\hbox{\rm const} \times
\hbox{\rm{coupling
\,const.}}\times k$. The linearly divergent piece of $\Pi(k)$ is
removed by
the mass counter term. Writing $\Pi(k)=Ak$ we have
\be
\Sigma(\Pi(p))=\int{d^3p\over(2\pi)^3}{1\over \bfp^2+Ap}
\approx\Sigma -A\int{d^3p\over(2\pi)^3}{1\over \bfp^3}
=\Sigma -8AH.
\ee

The calculation of the $A_i$ propagator gives \cite{appelqpisarski}
-\cite{nadkarni}
\be
\Pi_T(k)={1\over64}[(11-2)N_c-2]g_3^2k,
\ee
where $11N_c$ comes from the $A_i$ loop (11 is actually
$10+(1+\xi)^2$ in an
arbitrary covariant gauge), $-2N_c$ from the $A_0$ and $-2$ from the
$\phi$ loop.
For the $A_0$ and $\phi$ propagators
\be
\Pi_0(k)= {N_c\over4}g_3^2k, \qquad \Pi_\phi(k)={C_F\over4}g_3^2k.
\ee

Calculating finally the symmetry factors of the sunset diagrams one
obtains
the following result for the various diagrams in Fig.~2 contributing
to $\delta m^2$ ($C_F=(N_c^2-1)/(2N_c), [T^aT^b(T^aT^b+T^bT^a]_{ij}=
(N_c^2-1)(N_c^2-2)/(4N_c^2)\delta_{ij}$):
\begin{eqnarray}
(a)=\fr32(b) &=&-{3(N_c^2-1)(N_c^2-2)\over8N_c^2}g_3^4 H, \nonumber\\
(c)&=& -4(N_c+1)\lambda_3^2H, \nonumber\\
(d)+(e)+(f)&=&\fr14 C_F(9N_c-2)g_3^4H,  \\
(g)&=& 2C_FN_c g_3^4H,  \nonumber\\
(h)&=& 4C_F(N_c+1)\lambda_3 g_3^2H,\nonumber
\end{eqnarray}
The sunset diagrams come with a $-$ sign from the expansion of
$\exp(-\Gamma)\sim\int\exp(-S)$.
The final $\delta m^2$ is the sum of these; specializing to $N_c=2$
one obtains
the result in eq.~(\ref{count}).

Similarly, the diagrams in Fig.~3 contributing to $\delta m_D^2$ are
(using $[F^cF^d(F^cF^d+F^dF^c)]_{ab}=\fr32 N_c^2\delta_{ab}$,
$\tr T^aT^c(T^cT^b+T^bT^c)=(N_c^2-2)/(4N_c)\delta_{ab}$):
\begin{eqnarray}
(a)&=& -\fr94N_c^2 H, \nonumber\\
(b)&=& -(N_c^2+1)\lambda_A^2H,  \nonumber\\
(c)&=& -{N_c^2-2\over2N_c}g_3^4H,  \nonumber\\
(d)+(e)+(f)&=& \fr14 N_c(9N_c-2)H,  \\
(g)&=& 2N_c(N_c^2+1)\lambda_Ag_3^2H, \nonumber\\
(h)&=& 2C_Fg_3^4H.\nonumber
\end{eqnarray}
Now the sum (a)+(c)+(d)+(e)+(f)+(h)=0 and only the diagrams (b) and
(g)
containing the small induced coupling $\lambda_A$
survive in the result eq.(\ref{count}):
\be
\delta m_D^2 = (N_c^2+1)\lambda_A(2N_cg_3^2-\lambda_A).
\ee

\section{Appendix B}

In this Appendix we carry out the computation of the 2-loop effective
potential of the 3D effective theory defined by eq.(\ref{3daction}).
We give
further
the result obtained from this theory by integrating out the $A_0$
field
as well as the results also for the U(1) Higgs model both with and
without the $A_0$ field.

The relevant Feynman diagrams are shown in Fig.4, where the numbering
relates to Fig.~2. The diagrams come in two different variants (see
below), the sunset diagrams (a)-(c) and the closed figure-eight
diagrams
(d1)-(h4). The sunset diagrams are as in Fig.~2, the closed diagrams
are obtained by adding to the bubble diagrams (d)-(h) also the
corresponding
diagrams with a 4-vertex. Thus one obtains the figure-eight diagrams
in
Fig.~4: they did not contribute to the logarithmic divergence but
they
do contribute to the potential.

Landau gauge and dimensional regularization with $d=3-2\epsilon$
will be used throughout. The $\epsilon$ dependence is shown, if
needed,
only in connection of perturbative momentum space integrals.

\subsection{The scalar O(N) theory}
To show the technique \cite{jackiw}
and to carry out the renormalization in some detail,
consider first a 3d O(N) symmetric scalar
model with Lagrangian ($a=1,..,N$)
\be
{\cal L}= \fr12(\partial_i\phi_a)^2+\fr12m_0^2\phi_a^2+\fr14
\lambda(\phi_a^2)^2.
\ee
The dimensionalities of $\phi^2,\lambda,V$ are $\mu^{1-2\epsilon},
\mu^{1+2\epsilon},\mu^{3-2\epsilon}$, respectively.
As discussed earlier, there are only two divergent diagrams and one
only
has to introduce a mass counter term by writing $m_0^2=m^2_R+\delta
m^2$,
where $m_R^2$ is the renormalized mass.
After shifting $\phi_1\to\phi_0+\phi_1$ and neglecting
terms linear in $\phi_1$ the Lagrangian becomes
\begin{eqnarray}
{\cal L}&=&\fr12m_0^2\phi_0^2+\fr14\lambda\phi_0^4+
\fr12(\partial_i\phi_a)^2+
\fr12 m_1^2\phi_1^2+\fr12 m_2^2(\phi_2^2+...\phi_N^2)+ \\
&{\,}&+\lambda\phi_0\phi_1\phi_a^2+\fr14\lambda(\phi_a^2)^2\equiv
V_\rmi{tree}(\phi_0)+{\cal L}_0+{\cal L}_1,
\end{eqnarray}
where $m_1^2=m_0^2+3\lambda\phi_0^2,\,m_2^2=m_0^2+\lambda\phi_0^2$.
The
2-loop effective potential is then calculated from the functional
integral
\begin{eqnarray}
\exp[-{1\over\hbar}{\cal V}_3V_\rmi{eff}(\phi_0)]
&=&\exp[-{1\over\hbar}{\cal V}_3V_\rmi{tree}(\phi_0)]
\cdot \label{2loopformula0}\\ &&\cdot
\int {\cal D}\phi\exp\biggl[-{1\over\hbar}\int d^3x{\cal L}_0\biggr]
\langle\exp\biggl(-{1\over\hbar}\int d^3x{\cal L}_1\biggr) \rangle,
\nonumber
\end{eqnarray}
where ${\cal V}_3$ is the three-volume and where we have introduced a
formal expansion parameter $1/\hbar$. Rescaling the variable of
integration $\phi_a\to\phi_a\sqrt\hbar$ gives
\begin{eqnarray}
V_\rmi{eff}(\phi_0)&=&V_\rmi{tree}(\phi_0)-{\hbar\over{\cal V}_3}
\log
\int {\cal D}\phi\exp\biggl[-\int d^3x{\cal L}_0\biggr] -
\label{2loopformula}\\
&&-{\hbar\over{\cal V}_3}\langle\exp[-\int d^3x
(\lambda\phi_0\phi_1\phi_a^2
\sqrt\hbar+\fr14\lambda\phi_a^2\phi_b^2\hbar)]\rangle.  \nonumber
\end{eqnarray}
In the course of the rescaling $m_0^2$ becomes $m_0^2=m^2_R+\hbar^2
\delta m^2$; one $\hbar$ comes from the expansion of $\exp[-{\cal
L}_1]$,
the other is the $\hbar$ in front of the last term in
eq.(\ref{2loopformula}).

The Gaussian 1-loop integral is carried out in $\bfk$-space using
\be
\mu^{2\epsilon}\int{d^dk\over(2\pi)^d}\log(\bfk^2+m^2)
=-{1\over6\pi}m^3
\ee
with the result
\be
V_{(1)}(\phi_0)=-{\hbar\over12\pi}\mu^{-2\epsilon}[m_1^3+(N-1)m_2^3].
\label{(1)}
\ee
It is essential that we here can write $m_1^3=
[m^2_R+\hbar^2\delta m^2+3\lambda\phi_0^2]^{3/2} =
[m^2_R+3\lambda\phi_0^2]^{3/2}+{\cal O}(\hbar^2)$ so that the
$\delta m^2$ term only gives an $\hbar^3$ correction to the
potential,
which to 2 loops is calculated to order $\hbar^2$.

For the 2-loop contribution one first has to compute the expectation
value
of
\be
-\int d^3x{\cal L}_1=-\int d^3x
\hbar\fr14\lambda[\phi_1^2(x)+...+\phi_N^2(x)]
[\phi_1^2(x)+...+\phi_N^2(x)].
\ee
Each pairwise contraction gives (neglecting a common factor of ${\cal
V}_3$)
\be
I(m)=\mu^{-2\epsilon}\int{d^dk\over(2\pi)^d}{1\over \bfk^2+m^2}
=-{m\over4\pi}.
\ee
There are four different types of contractions corresponding to the
figure-eight diagram (h4). The first type is
$\langle \phi_1\phi_1\phi_1\phi_1 \rangle$, gives $3m_1^2/(4\pi)^2$
and appears only once. The second type
$\langle \phi_2\phi_2\phi_2\phi_2 \rangle$ appears $N-1$ times each
giving
$3m_2^2/(4\pi)^2$. The third type $\langle \phi_1\phi_1\phi_2\phi_2
\rangle$
gives $m_1m_2/(4\pi)^2$ and appears $2(N-1)$ times and the fourth
type
$\langle \phi_2\phi_2\phi_3\phi_3 \rangle$ appears $(N-1)(N-2)$ times
each giving $m_2^2/(4\pi)^2$. The total contribution to the potential
from the scalar figure-eight diagram, taking into account the sign of
$V_\rmi{eff}$ in eq.(\ref{2loopformula}), is
\be
V_{(2)}^{(h4)}(\phi_0)={\lambda\mu^{-4\epsilon}\hbar^2\over64\pi^2}
[3m_1^2+2(N-1)m_1m_2+(N^2-1)m_2^2]. \label{(h4)}
\ee

The second contribution to the 2-loop potential comes from the last
term
in eq.~(\ref{2loopformula}) and requires the calculation of the
vacuum
expectation value of
\be
\fr12 \hbar\lambda^2\int d^3x\int
d^3y\,\phi_0\phi_1(x)\phi_a(x)\phi_a(x)\cdot
\phi_0\phi_1(y)\phi_a(y)\phi_a(y).
\ee
This contains automatically the factor $\phi_0^2$ and corresponds to
the
sunset diagram (c). There are two different types of contractions,
$\langle\phi_1(x)\phi_1(x)\phi_1(x)\phi_1(y)\phi_1(y)\phi_1(y)\rangle$
appears once and gives $6H(m_1,m_1,m_1)$, where $H$ is the
dimensionless sunset function defined earlier, and
$\langle\phi_1(x)\phi_2(x)\phi_2(x)\phi_1(y)\phi_2(y)\phi_2(y)\rangle$
appears $N-1$ times each giving $2H(m_1,m_2,m_2)$. The total
contribution from the scalar sunset diagram is
\be
V_{(2)}^{(c)}(\phi_0)=-\lambda^2\mu^{-4\epsilon}\hbar^2\phi_0^2
[3H(m_1,m_1,m_1)+(N-1)H(m_1,m_2,m_2)]. \label{(c)}
\ee
Before renormalization the total 2-loop potential is the sum of
$V_\rmi{tree}$ and eqs.(\ref{(1)},\ref{(h4)},\ref{(c)}).


To carry out renormalization we insert
$H(m_1,m_2,m_2)=[1/(4\epsilon)+\log[\mu/(m_1+2m_2)]+1/2]/(16\pi^2)$
and find the following
coefficient of the $\fr12\phi_0^2$ term in the 2-loop potential:
\be
m^2_R+\hbar^2\delta
m^2-{6\lambda^2\mu^{-4\epsilon}\hbar^2\over16\pi^2}
\biggl({2+N\over3}{1\over4\epsilon}+\log{\mu\over3m_1}+{N-1\over3}
\log{\mu\over m_1+2m_2}+{N+2\over6}\biggr).
\ee
renormalization is carried out by choosing the counter term
\be
\delta m^2={2(2+N)\lambda^2\mu^{-4\epsilon}\over16\pi^2}
{1\over4\epsilon}.
\ee
Since the bare mass term $m_R^2+\hbar^2\delta m^2$ has to be
independent of $\mu$, this counter term implies that
\be
\mu{\partial m_R^2 \over\partial\mu}=-{f_{2m}\over16\pi^2},
\qquad f_{2m}=-2(N+2)\lambda^2. \label{mudependence}
\ee
The 2-loop potential in the final renormalized form then is
\begin{eqnarray}
V_\rmi{eff}(\phi_0)&=& \fr12 m_R^2(\mu) \phi_0^2 + \fr14
\lambda\phi_0^4-
\nonumber\\
&&-{\hbar\over12\pi}[m_1^3+(N-1)m_2^3]+ {\cal O}(\hbar^3) +
\label{Vscalar2loop}\\
&&+{\lambda\hbar^2\over64\pi^2}[3m_1^2+2(N-1)m_1m_2+(N^2-1)m_2^2]-
\nonumber\\
&&-3{\lambda^2\hbar^2\over16\pi^2} \biggl(
\log{\mu\over3m_1}+{N-1\over3}\log{\mu\over m_1+2m_2}+{N+2\over6}
\biggr)\phi_0^2
 + {\cal O}(\hbar^4).\nonumber
\end{eqnarray}
where $m_1^2=m_R^2(\mu)+3\lambda\phi_0^2$,
$m_2^2=m_R^2(\mu)+\lambda\phi_0^2$. In view of eq.(
\ref{mudependence}) one also explicitly has
\be
\mu{\partial V_\rmi{eff}(\phi_0) \over\partial\mu}= 0.
\ee

\subsection{SU(2)-Higgs theory}

The problem can be separated in two parts: computation of the
momentum integrals with massive propagators and computation of the
colour and symmetry factors. We shall mainly discuss the former.

There are three more difficult non-figure-eight
diagrams in Fig.4: the vector loop-scalar (VVS) diagram (a), the
scalar loop vector (SSV) diagrams (e1),(f1) and the vector
loop-vector
(VVV) diagram (d1).
After performing the propagator contractions in the
Landau gauge the VVS diagram is
\be
(a)=D_{VVS}(M,M,m)=\int{d^dp\over(2\pi)^d}{d^dq\over(2\pi)^d}
{(1-2\epsilon)\bfp^2\bfq^2+(\bfp\cdot\bfq)^2 \over
\bfp^2(\bfp^2+M^2)\bfq^2(\bfq^2+M^2)
[(\bfp+\bfq)^2+m^2]}. \label{VVS}
\ee
Here $d=3-3\epsilon$.
This diagram as well as the others are evaluated using the following:

1. Write for each combination
\be
{1\over \bfp^2(\bfp^2+m^2)}={1\over m^2} \biggl({1\over \bfp^2}-
{1\over \bfp^2+m^2} \biggr)\label{partialfraction}
\ee
and split each integral as a sum of zero-mass and massive terms.

2. If the integrand is of the form $f(\bfp^2)g(\bfq^2)
(\bfp\cdot\bfq)^n$, the integral vanishes for $n$ odd
and one can replace
\be
(\bfp\cdot\bfq)^2\to\fr1d\bfp^2\bfq^2,\quad
(\bfp\cdot\bfq)^4\to\frac{1}{d+2}\bfp^4\bfq^4.
\ee

d. Write $\bfp\cdot\bfq=\fr12[(\bfp+\bfq)^2-\bfp^2-\bfq^2]$ and add
and
subtract here mass terms so that you obtain same propagators as in
the
denominator.

4. Use that in dimensional regularization ($n,m,l$ are integer
positive
powers)
\be
\int{d^dp\over(2\pi)^d}{d^dq\over(2\pi)^d}{\bfp^n\bfq^m(\bfp+\bfq)^l
\over \bfp^2+m^2}=0
\ee
since the integral can be written as a sum of integrals from which
one is  scale-free and thus vanishing. Further
\be
I(m_1,m_2)=\int{d^dp\over(2\pi)^d}{d^dq\over(2\pi)^d}
{1\over (\bfp^2+m_1^2)
(\bfq^2+m_2^2)}={m_1m_2\over 16\pi^2},\label{defofi}
\ee
\be
L(m_1,m_2)=\int{d^dp\over(2\pi)^d}{d^dq\over(2\pi)^d}
{(\bfp\cdot\bfq)^2\over \bfp^2(\bfp^2+m_1^2)\bfq^2
(\bfq^2+m_2^2)}=\fr13{m_1m_2\over 16\pi^2},\label{defofl}
\ee
and
\begin{eqnarray}
&&\mu^{4\epsilon}\int {d^d p\over (2\pi)^d}{d^d k\over (2\pi)^d}
{1\over
(\bfp^2+m_1^2)(\bfk^2+m_2^2)(|\bfp+\bfk|^2+m_3^2)}=\label{defh}\\
&&=H(m_1,m_2,m_3)={1\over16\pi^2}\biggl({1\over4\epsilon} +
\log\frac{\mu}{m_1 + m_2 + m_3} +\fr12\biggr).
\nonumber
\end{eqnarray}
The VVS diagram in eq.(\ref{VVS}) then is, using rule 1.,
\begin{eqnarray}
(a)=D_{VVS}(M,M,m)&=&(1-2\epsilon)H(m,M,M)\nonumber \\
&&+{1\over M^4}[I(0,0,m)-2I(0,M,m)+I(M,M,m)],
\end{eqnarray}
where
\be
I(m_1,m_2,m_3)=\int{d^dp\over(2\pi)^d}{d^dq\over(2\pi)^d}
{(\bfp\cdot\bfq)^2 \over (\bfp^2+m_1^2)(\bfq^2+m_2^2)
[(\bfp+\bfq)^2+m_3^2]}.\label{defofimi}
\ee
In view of eq.~(\ref{defh}) the $-2\epsilon$ term here simply gives
the constant -1/2.
Calling the denominator in (\ref{defofimi}) $D_1D_2D_3$,
$D_1=\bfp^2+m_1^2$, etc., the numerator, using rule 4.,
becomes $[D_3^2+D_1^2+D_2^2+2D_1D_2-2D_2D_3-2D_1D_3-
2(m_3^2-m_2^2-m_1^2)(D_3-D_2-D_1)+ (m_3^2-m_2^2-m_1^2)^2]/4$. One
obtains
known integrals and
\be
\int{d^dp\over(2\pi)^d}{d^dq\over(2\pi)^d}{D_3\over D_1D_2}=
(m_3^2-m_2^2-m_1^2)m_1m_2
\ee
and permutations thereof. The sum is
\begin{eqnarray}
I(m_1,m_2,m_3)&=&\fr14\biggl[ (m_3^2-m_2^2-m_1^2)^2H(m_1,m_2,m_3)+
\nonumber\\ &&
+2(m_3^2-m_2^2-m_1^2)m_3(m_1+m_2)-\nonumber\\
&&-(m_3^2-m_2^2-m_1^2)m_1m_2
+(m_1^2-m_2^2-m_3^2)m_3m_2+\label{iii}\\
&&+(m_2^2-m_1^2-m_3^2)m_3m_1\biggr],\nonumber
\end{eqnarray}
which leads to the final expression for the momentum integral for
diagram (a):
\begin{eqnarray}
(a)=D_{VVS}(M,M,m)&=&
\biggl\{2H(M,M,m)-{1\over2}H(M,m,0)+
\nonumber\\ &{\,}&
+{m^2\over M^2}\bigl[H(M,m,0)-H(M,M,m) \bigr]+ \\
&{\,}& +{m^4\over4M^4} \bigl[H(m,0,0)+H(M,M,m)-2H(M,m,0)
\bigr]-\nonumber\\ &{\,}&
-{m\over2M}-{m^2\over4M^2} \biggr\}\nonumber
\end{eqnarray}

The second more complicated diagram is the scalar loop-vector
diagram (SSV):
\be
(e1),(f1)=D_{SSV}(m_1,m_2,M)=\int{d^dp\over(2\pi)^d}{d^dq\over(2\pi)^d}
{4[\bfp^2\bfq^2-(\bfp\cdot\bfq)^2] \over
(\bfp^2+m_1^2)\bfq^2(\bfq^2+m_V^2)
[(\bfp+\bfq)^2+m_2^2]}.
\ee
This equals
\be
D_{SSV}(m_1,m_2,M)=\int{d^dp\over(2\pi)^d}{d^dq\over(2\pi)^d}
\biggl[ {4\over D_2D_3}-{4m_1^2\over D_1D_2D_3}\biggr]
-{4\over M^2}[I(m_1,0,m_2)-I(m_1,M,m_2)]
\ee
and one can directly use the integral in eq.(\ref{iii}) with the
result
\begin{eqnarray}
D_{SSV}(m_1,m_2,M)&=&(M^2-2m_1^2-2m_2^2)H(M,m_1,m_2)+\nonumber\\
&&+{(m_1^2-m_2^2)^2\over M^2}[H(M,m_1,m_2)-H(0,m_1,m_2)]+
\label{SSV}\\
&&+{1\over M}\{(m_1+m_2)[M^2+(m_1-m_2)^2]-Mm_1m_2\}. \nonumber
\end{eqnarray}

The most complicated diagram is the vector loop-vector diagram (VVV):
\begin{eqnarray}
&&(d1)=D_{VVV}(m,m,m)= \label{VVV}\\
&&\int{d^dp\over(2\pi)^d}{d^dq\over(2\pi)^d}
{[(\bfp\cdot\bfq)^2-\bfp^2\bfq^2] [2\bfp^4+4\bfp^2\bfp\cdot\bfq
+(\bfp\cdot\bfq)^2+5\bfp^2\bfq^2+4\bfp\cdot\bfq \bfq^2+2\bfq^4]
\over \bfp^2(\bfp^2+m^2)\bfq^2(\bfq^2+m^2)(\bfp+\bfq)^2
[(\bfp+\bfq)^2+m^2]}\nonumber\\
&&-2\epsilon\int{d^dp\over(2\pi)^d}{d^dq\over(2\pi)^d}
{[(\bfp\cdot\bfq)^2-\bfp^2\bfq^2] [\bfp^4+\bfq^4+
2(\bfp^2+\bfq^2)\bfp\cdot\bfq+d\bfp^2\bfq^2]
\over \bfp^2(\bfp^2+m^2)\bfq^2(\bfq^2+m^2)(\bfp+\bfq)^2
[(\bfp+\bfq)^2+m^2]}.\nonumber
\end{eqnarray}
Denoting the integral without the $\bfp^2,\bfq^2$ and $(\bfp+\bfq)^2$
factors in the denominator by $(m,m,m)$, one obtains, by using
eq.(\ref{partialfraction})
\be
D_{VVV}(m,m,m)=-{1\over m^6}\biggl[(m,m,m)-3(m,m,0)+3(m,0,0)-(0,0,0)
\biggr].
\ee
The reduction techniques explained earlier lead to
\begin{eqnarray}
(m,m,m)&=&-{63\over16}m^8H(m,m,m)+{183\over16}m^8+\fr98
m^8,\nonumber\\
(m,m,0)&=&{10\over3}m^8,\\
(m,0,0)&=&{1\over16}m^8H(m,0,0), \nonumber
\end{eqnarray}
where the last term in $(m,m,m)$ arises from the $-2\epsilon$ term in
(\ref{VVV}). Adding up gives
\be
D_{VVV}(m,m,m)=m^2\biggl[{63\over16}H(m,m,m)-{3\over16}H(m,0,0)-
{41\over16} \biggr].
\ee

The figure-eight diagrams lead to integrals of the type in
eqs.(\ref{defofi}-\ref{defofl}) and are easy to evaluate.
Including the charge and symmetry factors
leads to the final result for the diagrams ($\times16\pi^2$) in
Fig.4:
\begin{eqnarray}
(a)&=&-{3\over16}g_3^4\phi^2
\biggl\{2H(m_1,m_T,m_T)-{1\over2}H(m_1,m_T,0)+
\nonumber\\ &{\,}&
+{m_1^2\over m_T^2}\bigl[H(m_1,m_T,0)-H(m_1,m_T,m_T) \bigr]+
\label{diagram_a}\\
&{\,}& +{m_1^4\over4m_T^4}
\bigl[H(m_1,0,0)+H(m_1,m_T,m_T)-2H(m_1,m_T,0)
\bigr]- \nonumber\\ &{\,}&
-{m_1\over2m_T}-{m_1^2\over4m_T^2} \biggr\}, \nonumber\\
(b)&=&-{3\over16}g_3^4\phi^2H(m_1,m_L,m_L), \\
(c)&=&-3\lambda_3^2\phi^2 \bigl[H(m_1,m_1,m_1)+H(m_1,m_2,m_2) \bigr],
\\
(d1)&=&
2g_3^2\bigl[{63\over16}m_T^2H(m_T,m_T,m_T)-{3\over16}m_T^2H(m_T,0,0)
-{41\over16}m_T^2 \bigr], \\
({\rm ghost}) &=& 2g_3^2{3\over8}m_T^2H(m_T,0,0), \\
(d2)&=&4g_3^2m_T^2, \\
(e1)&=& -{3\over2}g_3^2\bigl[
(m_T^2-4m_L^2)H(m_T,m_L,m_L)+2m_Tm_L-m_L^2
\bigr],  \\
(e2) &=& 6g_3^2m_Tm_L ,\\
(f1) &=& -{3\over8}g_3^2 \biggl[ (m_T^2-2m_1^2-2m_2^2)H(m_1,m_2,m_T)
+(m_T^2-4m_2^2)H(m_2,m_2,m_T)+\nonumber\\
&{\,}&\quad +{(m_1^2-m_2^2)^2\over
m_T^2}[H(m_1,m_2,m_T)-H(m_1,m_2,0)]+ \\
&{\,}&\quad+{(m_1^2-m_2^2)(m_1-m_2)\over
m_T}+m_T(m_1+3m_2)-m_1m_2-m_2^2
\biggr], \nonumber\\
(f2)&=& {3\over4}g_3^2m_T(m_1+3m_2), \\
(g3)&=& {15\over4}\lambda_A m_L^2,\\
(g4)&=& {3\over8}g_3^2m_L(m_1+3m_2), \\
(h4)&=& {3\over4}\lambda_3(m_1^2+2m_1m_2+5m_2^2).\label{diagram_h4}
\end{eqnarray}
It may also be useful to quote the sum of all the non-Abelian
diagrams
\begin{eqnarray}
&&(d1)+({\rm ghost})+(d2)+(e1)+(e2)=\nonumber\\
&=&g_3^2m_T^2\biggl[{63\over8}H(m_T,m_T,m_T)+\fr92H(m_T,m_L,m_L)+
\fr38H(m_T,0,0)+{3\over8}\biggr]+\label{gluondiags}\\
&&+3g_3^2m_Lm_T+6g_3^2m_3^2H(m_T,m_L,m_L).\nonumber
\end{eqnarray}

\subsection{U(1)-Higgs theory}
In this appendix we consider 4d U(1) Higgs theory defined by
Lagrangian
\be
L={1\over4} F_{\mu\nu}F_{\mu\nu} +
(D_{\mu}\phi)^\dagger(D_{\mu}\phi) -\fr12 m^2\phi^\dagger\phi
+\lambda(\phi^\dagger\phi)^2.
\label{4dlagrU(1)}
\ee

The dimensionally reduced theory is
\begin{eqnarray}
&&S_{\rmi{eff}}[A_i(\bfx),A_0(\bfx),\phi(\bfx)]
= \int d^3x \biggl\{{1\over4} F_{ij}F_{ij} +
\fr12 (D_iA_0)(D_iA_0) + (D_i\phi)^\dagger(D_i\phi)+
\nonumber
\\&&
\fr12 m_D^2 A_0A_0 + \frac{1}{4}\lambda_A (A_0A_0)^2
+ m_3^2\phi^\dagger\phi
+\lambda_3(\phi^\dagger\phi)^2
+ h_3 A_0^aA_0^a \phi^\dagger\phi \biggr\}.
\label{3dactionU(1)}
\end{eqnarray}
The one-loop relations relations between 3d and 4d couplings and
fields in the Landau gauge are:
\be
g_3^2 = g^2(\mu)T[1 - \frac{g^2 L_s}{3(4 \pi)^2}],
\label{g3u1}
\ee
\be
\lambda_3 = T[\lambda(\mu) - \frac{L_s}{(4 \pi)^2}(3g^4
-6\lambda g^2 + 10 \lambda^2)+\frac{2}{(4
\pi)^2}g^4],
\label{lambda3u1}
\ee
\be
h_3 = g^2(\mu)T[1 - \frac{g^2 L_s}{3(4 \pi)^2}
+ \frac{4 g^2}{3(4 \pi)^2}+\frac{8 \lambda}{(4 \pi)^2}]=
\label{h3u1}
\ee
\[\frac{1}{4}g_3^2[1 + \frac{4 g^2}{3(4 \pi)^2}+\frac{8 \lambda}{(4
\pi)^2}],
\]
\be
\lambda_A = \frac{g^4(\mu) T}{24 \pi^2},
\label{lambdaAu1}
\ee
\be
m_D^2 = \frac{1}{3} g^2(\mu) T^2,
\label{mDu1}
\ee
\be
m_3^2 = [{1\over4} g^2(\mu) + \fr13 \lambda(\mu)]T^2 - \fr12
m^2(\mu)[1 +
\frac{L_s}{(4 \pi)^2}(-3 g^2 + 4 \lambda)],
\label{m3u1}
\ee
\be
\phi^{3d} = \frac{1}{\sqrt{T}}\phi(1-\frac{3g^2}{2(4
\pi)^2}\frac{1}{\epsilon_B}),
\ee
\be
A_0^{3d} = \frac{1}{\sqrt{T}}A_0[1+\frac{g^2}{6(4 \pi)^2}
(\frac{1}{\epsilon_B}+2)],
\ee
\be
A_i^{3d} = \frac{1}{\sqrt{T}}A_i[1+\frac{g^2}{6(4 \pi)^2}
\frac{1}{\epsilon_B})].
\ee

As in the main body of the paper, we present the 2-loop potential
with the use of simplified relations between couplings, namely
\be
h_3 = g_3^2,~\lambda_A=0
\ee
The field dependent masses are:
\be
m_T=g_3\phi,~ m_L^2=m_D^2+m_T^2=\fr13 g^2T^2+m_T^2~,
\ee
\[
m_1^2=m_3^2(\mu_3)+3\lambda_3\phi^2,~
m_2^2=m_3^2(\mu_3)+\lambda_3\phi^2.
\]
The 1-loop potential in 3d is
\be
V(\phi)=\fr12
m_3^2(\mu_3)+\fr14\lambda_3
\phi^4-{T\over12\pi}(2m_T^3+m_L^3+m_1^3+m_2^3).
\ee
In the calculation of the 2-loop potential in 3d
the main difference with respect to SU(2) is that the non-Abelian
diagrams in eq.~(\ref{gluondiags}) do not appear. Secondly, since
there is only one Higgs and one Goldstone mode, the diagrams with
two different Goldstone modes are missing. The color factors
clearly also are different. The final result for the diagrams
($\times16\pi^2$ for non-$H$ diagrams) is
\begin{eqnarray}
(a)&=&-g^4\phi^2D_{VVS}(m_1,m_T,m_T), \nonumber\\
(b)&=&-g^4\phi^2H(m_1,m_L,m_L), \nonumber\\
(c)&=&-\lambda^2\phi^2[3H(m_1,m_1,m_1)+H(m_1,m_2,m_2)], \nonumber\\
(f1)&=&-\fr12g^2D_{SSV}(m_1,m_2,m_T),\label{u1diags} \\
(f2)+(g4)&=&\fr12g^2(m_L+2m_T)(m_1+m_2), \nonumber\\
(h4)&=&\fr14\lambda(3m_1^2+2m_1m_2+3m_2^2). \nonumber
\end{eqnarray}
Again the coefficient of $\fr12\phi^2\log(\mu_3)$ in the sum gives
the
value of $f_{2m}/16\pi^2$ in the 3d running of the scalar mass
$m_3^2(mu_3)$,
\be
m_3^2(\mu_3)={f_{2m}\over16\pi^2}\log{\Lambda_m\over\mu_3},
\ee
where
\be
f_{2m}=-6g_3^4+8\lambda_3
g_3^2-8\lambda_3^2=-6g_3^4\biggl(1-{2m_H^2\over3m_W^2}
+{m_H^4\over3m_W^4}\biggr)<0. \label{f2mforu1}
\ee
with $\lambda_3 = g_3^2m_H^2/(2m_W^2)$. The relation between the
temperature and the renormalization invariant scale $\Lambda_m$ can
be found as has been described in the Section. 6.

\subsection{SU(2)-Higgs theory integrated over $A_0$}
In view of the large value of $m_D$ one may further simplify the
effective theory by entirely integrating out the $A_0$ field, the
effective action then becomes $S_\rmi{eff}[A_i^a,\phi_k]$. The
integration can be carried out explicitly, but it is rather
illuminating
to derive the result by taking the large $m_D$ limit of the 2-loop
effective potential.

The relevant diagrams in eqs.~(\ref{diagram_a}-\ref{diagram_h4})
are those containing $m_L=\sqrt{m_D^2+m_T^2}$. Expanding them
to order $1/m_D$ and neglecting a constant term $\sim
m_D^2[\log(\mu/2m_D)+1/4]$ one obtains for the relevant diagrams (in
evident notation)
\begin{eqnarray}
(A_0A_0\phi)&=-&{3\over16}g^4\phi^2(\log{\mu\over 2m_D}+\fr12)+
\fr38g^2m_T^2{m_1\over m_D}+\dots,\nonumber\\
(A_0A_0A_i)+(A_0A_i)&=&\fr92 g^2m_T^2(\log{\mu\over
2m_D}+\fr12)-\fr34g^2m_T^2
+{g^2\over2m_D}m_T^3+\dots,\\
(A_0\phi)&=&\fr38g^2m_D(m_1+3m_2)+{3\over16}g^2m_T^2{m_1+3m_2\over
m_D}
+\dots . \nonumber
\end{eqnarray}
Expanding also
\be
m_L^3=m_D^3+\fr32m_Dm_T^2+\fr38{m_T^4\over m_D}+\dots\nonumber
\ee
gives the limit of the 2-loop potential as
\begin{eqnarray}
V(m_D\gg g^2)&=&\fr12\bigl[m^2(\mu)-{3g^2m_D\over16\pi}
+{15\over(16\pi)^2}g^4(\log{\mu\over2m_D}+\fr12)-{3g^4\over128\pi^2}\biggr]\phi^2\nonumber\\
&&+
\fr14\biggl(\lambda-{3g^4\over128\pi m_D}\biggr)\phi^4-
-{1\over12\pi}\bigl[6m_T^3+m_1^3+3m_2^3\bigr]+\label{mdlimit}\\
&&+{g^2\over32\pi^2m_D}m_T^3+{3\over128\pi^2}g^2m_D(m_1+3m_2)+
{9g^2\over256\pi^2m_D}m_T^2(m_1+m_2)+\nonumber\\
&&+{\rm 2-loop\,\,diagrams\,\,in\,\,
eqs.~(\ref{diagram_a}-\ref{diagram_h4})\,\,without\,\,A_0}.\nonumber
\end{eqnarray}
If we now introduce new parameters by the following equations
\begin{eqnarray}
\bar m_3^2(\mu_3)&=&m_3^2(\mu_3)-{3g_3^2m_D\over16\pi}+
{30\over(16\pi)^2}g_3^4(\log{\mu_3\over2m_D}+\frac{3}{10}),\nonumber\\
\bar g_3^2&=&g_3^2\biggl(1-{g_3^2\over24\pi
m_D}\biggr),\label{newparam}\\
\bar\lambda_3&=&\lambda_3-{3g_3^4\over128\pi m_D}, \nonumber
\end{eqnarray}
we see that the $m_D$-dependent terms in eq.~(\ref{mdlimit}) are
entirely cancelled. This is obviously so for the $\phi^2$ and
$\phi^4$
terms. The transformation of $g^2$ combines the $m_T^3=g^3\phi^3/8$
terms and inserting the transformation of $m^2(\mu)$ and $\lambda$ in
$m_1^3+3m_2^3$ cancels the two last terms in eq.~(\ref{mdlimit}). The
$g^4\log(\mu/2m_D)$ term simply removes the $A_0$ loop contributions
from $f_{2m}$ and is otherwise of higher order--as is also the last
term
in the transformation of $m_3^2(\mu_3)$.

The effective potential of the effective theory obtained by
integrating
over $A_0$ is thus
\begin{eqnarray}
V\rmi{eff}&=&\fr12\bar m_3^2(\mu_3)\phi^2+\fr14\bar\lambda_3\phi^4
-{1\over12\pi}\bigl[6\bar m_T^3+\bar m_1^3+3\bar m_2^3\bigr]+
\nonumber\\
&&+{\rm 2-loop\,\,diagrams\,\,in\,\,
eqs.~(\ref{diagram_a}-\ref{diagram_h4})\,\,without\,\,A_0},
\end{eqnarray}
where $\bar m_T=\fr12\bar g_3\phi$, $\bar m_1^2=\bar m_3^2(\mu_3)+3
\bar\lambda_3\phi^2$ and $\bar m_2^2=\bar
m_3^2(\mu)+\bar\lambda_3\phi^2$.
It is formally irrelevant whether one uses $g_3$ or $\bar g_3$ in the
2-loop diagrams; the difference is of higher order.

A direct diagrammatic derivation of eqs.(\ref{newparam}) is
straightforward:
the first requires the evaluation of the $A_0$ 1-loop and 2-loop
contribution to the $\phi$ propagator, second the evaluation of the
$k^2$ term in the $A_0$ loop contribution to the $A_i$ propagator
and the last the $A_0$ loop contribution to the self-coupling.

\subsection{U(1)-Higgs theory integrated over $A_0$}
The effective potential of the theory obtained from the U(1) theory
after integrating out the $A_0$ field is
\begin{eqnarray}
V\rmi{eff}&=&\fr12\bar m^2(\mu)\phi^2+\fr14\bar\lambda\phi^4
-{1\over12\pi}\bigl[2\bar m_T^3+\bar m_1^3+\bar m_2^3\bigr]+
\nonumber\\
&&+{1\over16\pi^2}\biggl\{-g^4\phi^2D_{VVS}(m_1,m_T,m_T)-
\fr12g^2D_{SSV}(m_1,m_2,m_T)\nonumber\\
&&-\lambda^2\phi^2[3\bar H(m_1,m_1,m_1)+\bar H(m_1,m_2,m_2)]+\\
&&+g^2m_T(m_1+m_2)+\fr14(3m_1^2+2m_1m_2+3m_2^2)\biggr\},\nonumber
\end{eqnarray}
where the couplings are related to those of the original theory by
\begin{eqnarray}
\bar m_3^2(\mu_3)&=&m_3^2(\mu_3)-{g_3^2m_D\over4\pi}-
{1\over8\pi^2}g_3^4(\log{\mu_3\over2m_D}+\fr12),\nonumber\\
\bar\lambda_3&=&\lambda_3-{g_3^4\over8\pi m_D}. \label{newparamu1}
\end{eqnarray}
Due to the absence of non-Abelian coupling there is no rescaling of
$g_3^2$.

\newpage
\begin{figure}
\psfig {file=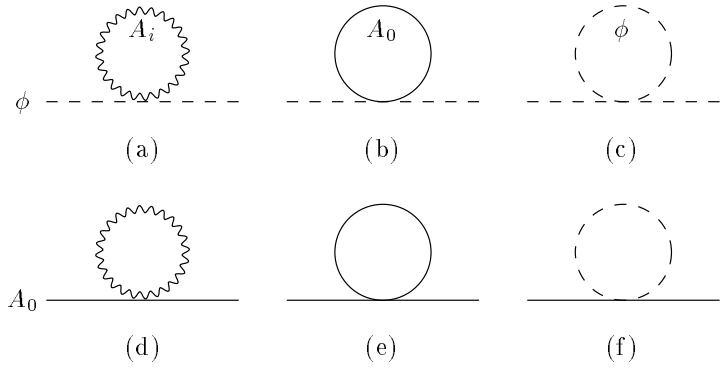,width=7.5in}
\vspace{-15cm}
\caption{The one-loop linear-divergent diagrams contributing to the
mass of the doublet and the triplet of scalar fields.}
\end{figure}
\begin{figure}
\psfig {file=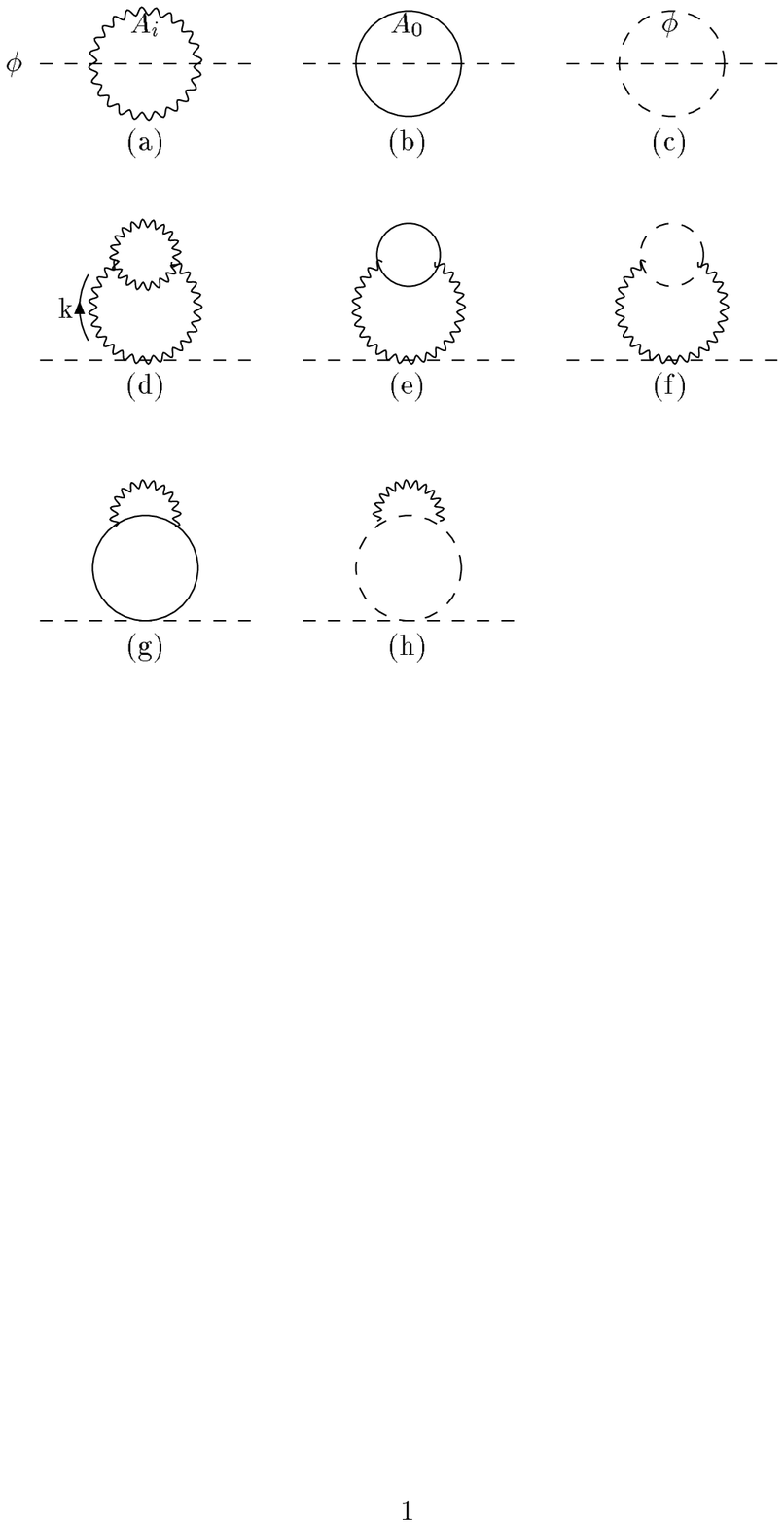,width=7.5in}
\vspace{-12cm}
\caption{The two-loop log-divergent diagrams contributing to the mass
of the doublet of scalar fields.}
\end{figure}
\begin{figure}
\psfig {file=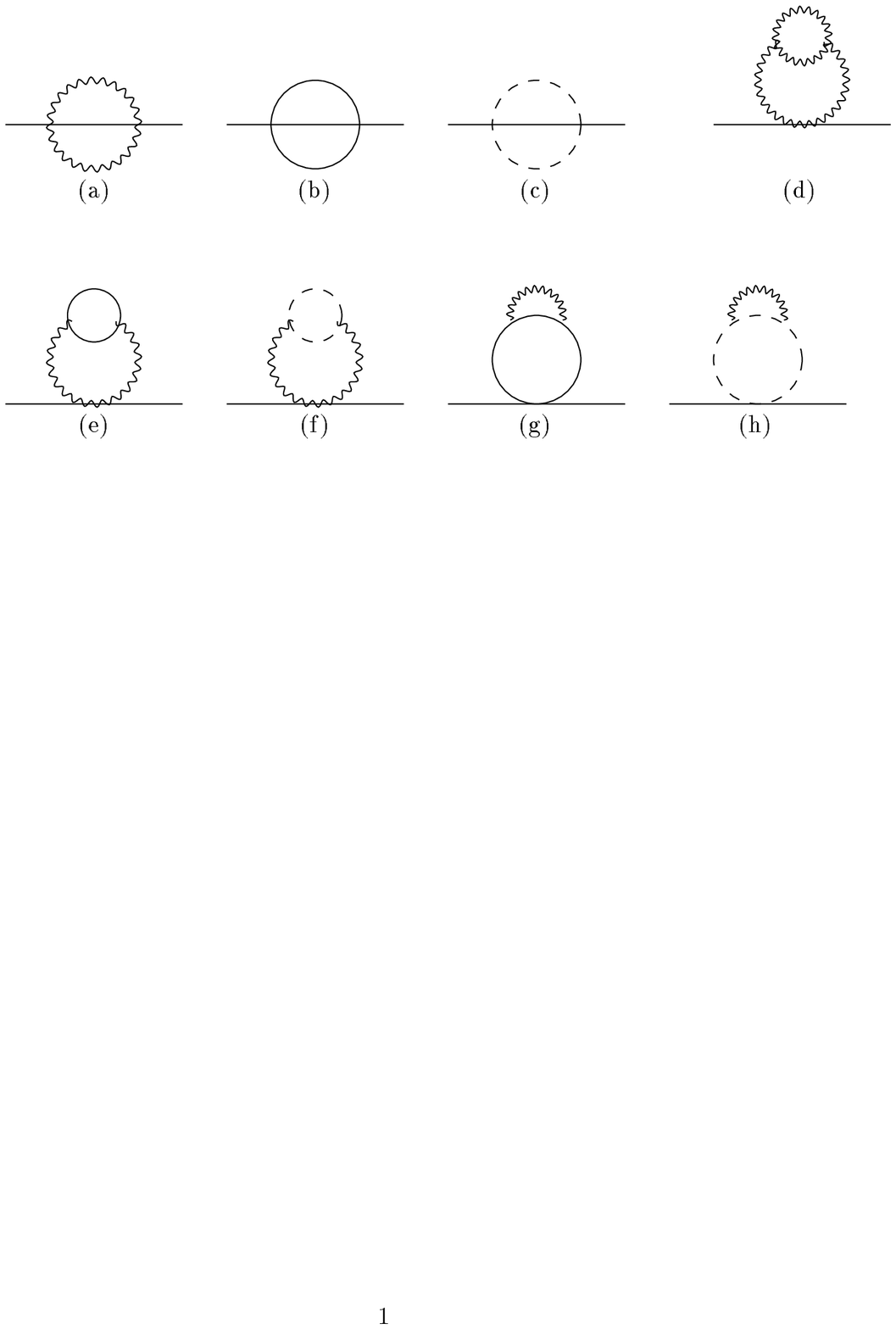,width=7.5in}
\vspace{-14cm}
\caption{The two-loop log-divergent diagrams contributing to the mass
of the triplet of scalar fields.}
\end{figure}
\begin{figure}
\psfig {file=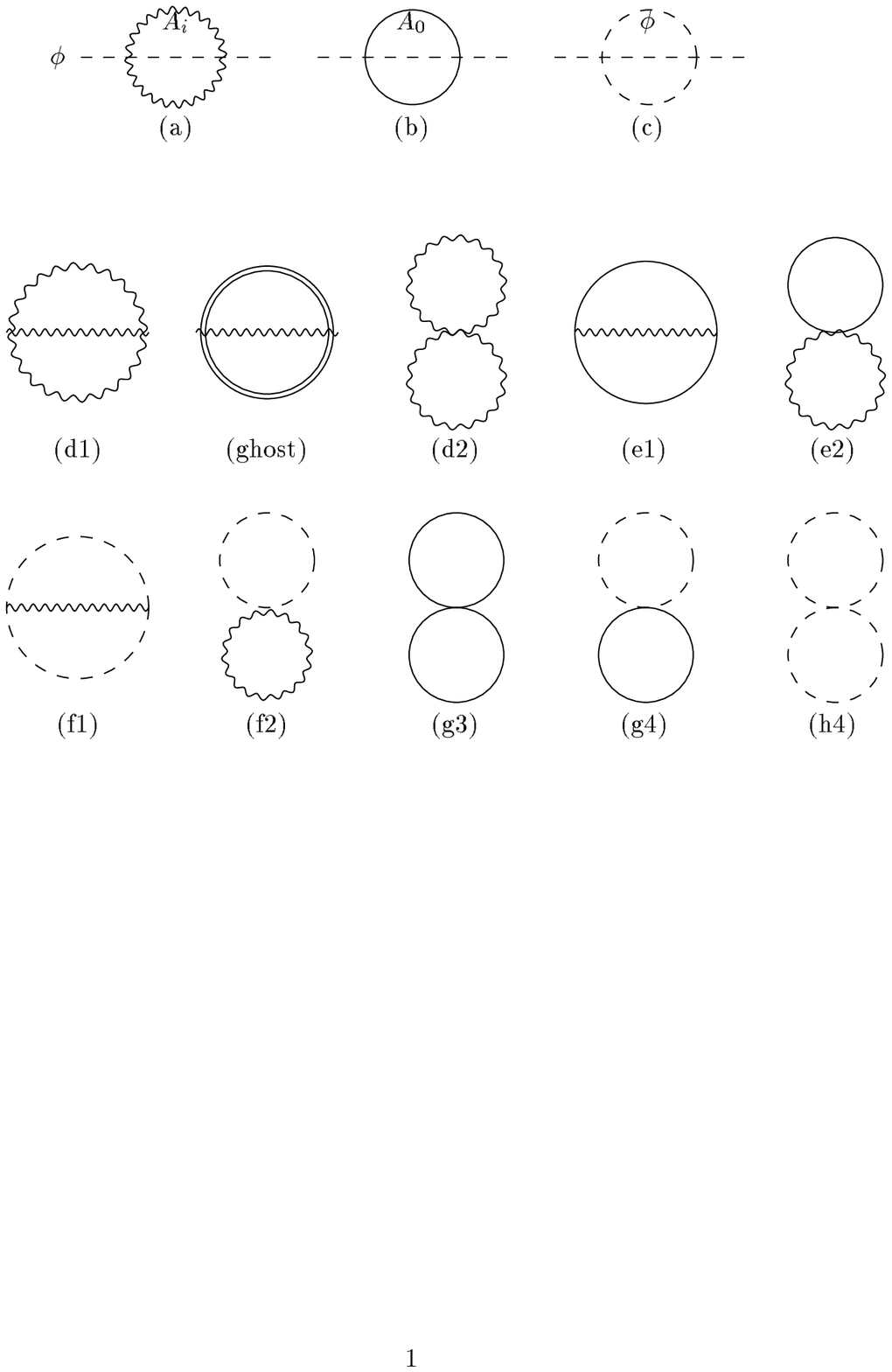,width=7.5in}
\vspace{-11cm}
\caption{The two-loop contributions to the effective potential.}
\end{figure}
\begin{figure}
\vspace{-4cm}
\psfig {file=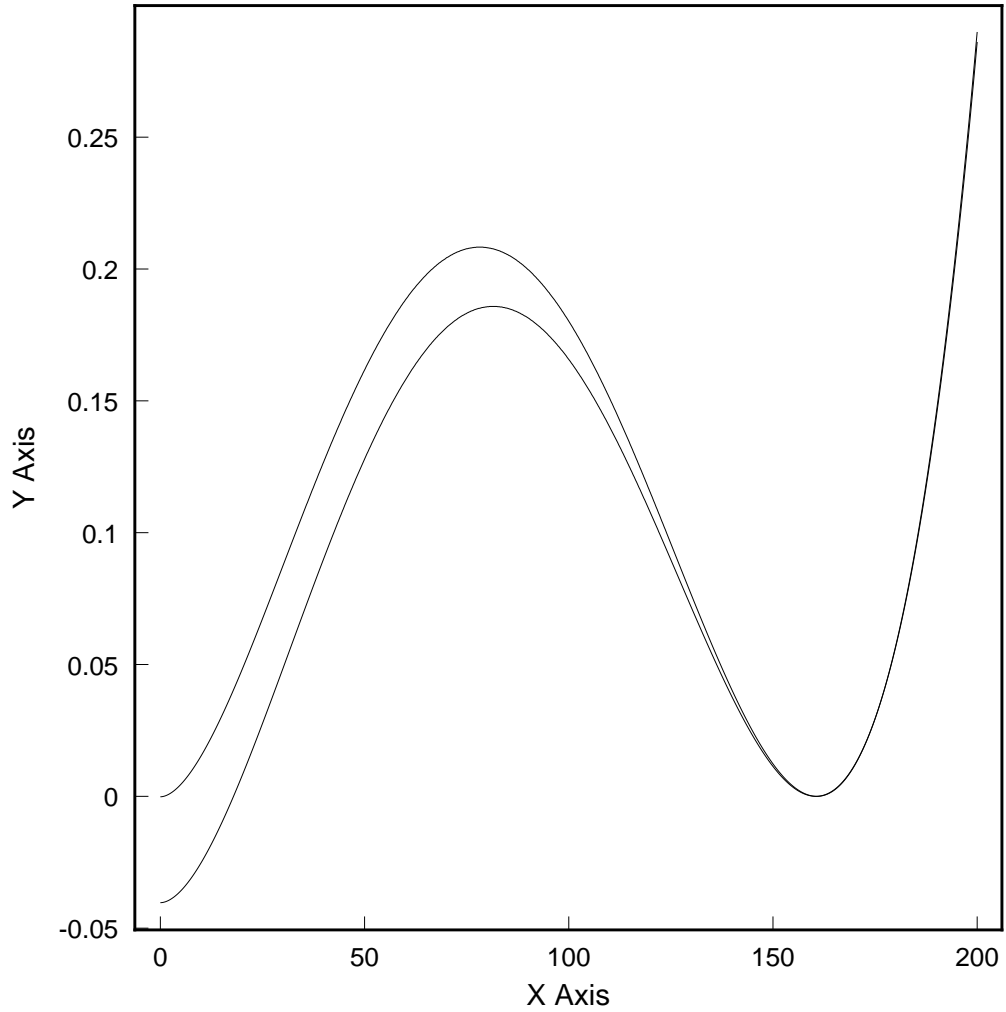,width=5.5in}
\caption{The two-loop (upper curve) and the one-loop (lower curve) rg
improved effective potentials at the 2-loop critical temperature for
$m_H = 35$ GeV. The x-axis is 4d scalar field, in GeV, the y
axis is a dimensionless 3d potential $\frac{12}{g_3^6}V(\phi)$.}
\end{figure}
\begin{figure}
\psfig {file=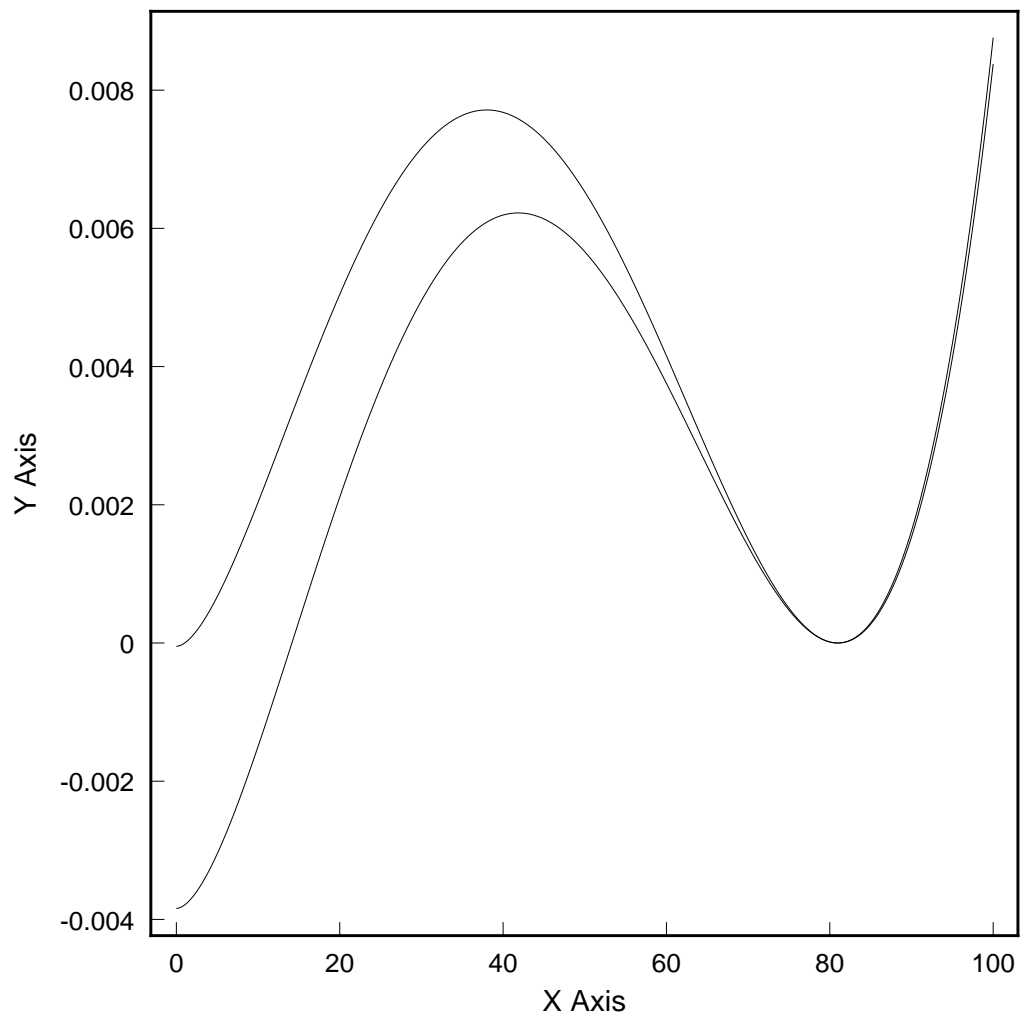,width=5.5in}
\caption{The same as in Fig. 5  for $m_H = 80 GeV$.}
\end{figure}
\end{document}